\documentclass[journal,onecolumn,11pt]{IEEEtran}

\usepackage{latexsym}
\usepackage{graphics}
\usepackage{times}
\usepackage{amssymb,amsmath,graphicx,amsfonts}
\usepackage{setspace}

\newtheorem{theorem}{Theorem}
\newtheorem{corollary}{Corollary}

\providecommand{\E}{{\sf E}} 
\providecommand{\Cov}{{\sf Cov}} \providecommand{\Pr}{{\sf Pr}}
\providecommand{\tr}{{\sf Tr}}

 \providecommand{\Av}{\mathbf{A}}
 \providecommand{\Bv}{\mathbf{B}}
 
 \providecommand{\Dv}{\mathbf{D}}

 \providecommand{\Iv}{\mathbf{I}}

 \providecommand{\Nv}{\mathbf{N}}
 \providecommand{\Mv}{\mathbf{M}}

 \providecommand{\Sv}{\mathbf{S}}
 
 \providecommand{\Uv}{\mathbf{U}}
 \providecommand{\Vv}{\mathbf{V}}

\providecommand{\xv}{\mathbf{x}} \providecommand{\Xv}{\mathbf{X}}
 \providecommand{\Yv}{\mathbf{Y}}
 \providecommand{\Zv}{\mathbf{Z}}

 \providecommand{\Rs}{{\mathbb R}}

\providecommand{\Zt}{\widetilde{\mathbf{Z}}}

\providecommand{\Nt}{\widetilde{\mathbf{N}}}

\providecommand{\T}{\textsf{T}}

\begin{document}
\doublespace

\title{A Vector Generalization of Costa's Entropy-Power\\[-2mm] Inequality with Applications}


\author{Ruoheng Liu, Tie Liu, H. Vincent Poor, and Shlomo Shamai (Shitz)%
\thanks{This research was supported by the United States National Science Foundation under Grants
CNS-06-25637 and CCF-07-28208, the European Commission in the
framework of the FP7 Network of Excellence in Wireless
Communications NEWCOM++, and the Israel Science Foundation. The
material in this paper was presented in part at the New Result
Session of the 2008 IEEE International Symposium on Information
Theory, Toronto, Ontario, Canada, July 2008.}%
\thanks{Ruoheng Liu and H. Vincent Poor are with the Department of Electrical Engineering,
Princeton University, Princeton, NJ 08544, USA. Email: {\{rliu,poor\}@princeton.edu}}%
\thanks{Tie Liu is with the Department of Electrical and Computer Engineering, Texas
A\&M University, College Station, TX 77843, USA. Email: tieliu@tamu.edu}%
\thanks{Shlomo Shamai (Shitz) is with the Department of Electrical Engineering,
Technion-Israel Institute of Technology, Technion City, Haifa 32000,
Israel. Email: sshlomo@ee.technion.ac.il}}

\maketitle

\begin{abstract}
This paper considers an entropy-power inequality (EPI) of Costa and presents a
natural vector generalization with a real positive semidefinite matrix
parameter. This new inequality is proved using a perturbation approach via a
fundamental relationship between the derivative of mutual information and the
minimum mean-square error (MMSE) estimate in linear vector Gaussian channels.
As an application, a new extremal entropy inequality is derived from the
generalized Costa EPI and then used to establish the secrecy capacity regions
of the degraded vector Gaussian broadcast channel with layered confidential
messages.
\end{abstract}

\begin{keywords}
Entropy-power inequality (EPI), extremal entropy inequality,
information-theoretic security, mutual information and minimum
mean-square error (MMSE) estimate, vector Gaussian broadcast channel
\end{keywords}

\section{Introduction}
In information theory, the entropy-power inequality (EPI) of Shannon
\cite{Sha-BSTJ48} and Stam \cite{Sta-IC59} has played key roles in
the solution of several canonical network communication problems.
Celebrated examples include Bergmans's solution \cite{Ber-IT73} to
the Gaussian broadcast channel problem, Leung-Yan-Cheong and
Hellman's solution \cite{Cheong:IT:78} to the Gaussian wire-tap
channel problem, Ozarow's solution \cite{Oza-BSTJ80} to the Gaussian
two-description problem, Oohama's solution \cite{Ooh-IT98} to the
quadratic Gaussian CEO problem, and more recently Weingarten,
Steinberg and Shamai's solution \cite{WSS-IT06} to the
multiple-input multiple-output Gaussian broadcast channel problem.

Let $\Xv$ and $\Zv$ be two independent random $n$-vectors with densities in
$\Rs^{n}$, where $\Rs$ denotes the set of real numbers. The classical EPI of
Shannon \cite{Sha-BSTJ48} and Stam \cite{Sta-IC59} can be written as
\begin{align}
\exp\left[\frac{2}{n}h(\Xv+\Zv)\right]\ge
\exp\left[\frac{2}{n}h(\Xv)\right]+\exp\left[\frac{2}{n}h(\Zv)\right]
\label{eq:epi}
\end{align}
where $h(\Xv)$ denotes the differential entropy of $\Xv$. The
equality holds if and only if $\Xv$ and $\Zv$ are Gaussian and with
proportional covariance matrices.

In network information theory, most applications focus on the special case of
\eqref{eq:epi} where one of the random vectors is fixed to be Gaussian. In this
setting, the classical EPI of Shannon and Stam can be further strengthened as
shown by Costa \cite{Cos-IT85}. Let $\Zv$ be a Gaussian random $n$-vector with
a positive definite covariance matrix, and let $a$ be a real scalar such that
$a \in [0,1]$. Costa's EPI \cite{Cos-IT85} can be written as
\begin{align}
\exp\left[\frac{2}{n}h(\Xv+\sqrt{a}\Zv)\right] & \geq
(1-a)\exp\left[\frac{2}{n}h(\Xv)\right]+ a
\exp\left[\frac{2}{n}h(\Xv+\Zv)\right] \label{eq:SC-epi}
\end{align}
for any random $n$-vector $\Xv$ independent of $\Zv$. The equality
holds if and only if $\Xv$ is also Gaussian and with a covariance
matrix proportional to that of $\Zv$'s.

Though not as widely known as the classical EPI of Shannon and Stam, Costa's
EPI has found useful applications in deriving capacity bounds for the Gaussian
interference channel \cite{Cos-IT85-IC} and the multiantenna flat-fading
channel \cite{LM-IT03}. The original proof of Costa's EPI provided in
\cite{Cos-IT85} was based on rather detailed calculations. Simplified proofs
based on a Fisher information inequality \cite{DCT-IT91} and a fundamental
relationship between the derivative of mutual information and minimum
mean-square error (MMSE) in linear Gaussian channels \cite{GSV-IT05} can be
found in \cite{Dem-IT89} and \cite{GSV-ISIT06}, respectively.

Note that Costa's EPI \eqref{eq:SC-epi} provides a strong
relationship among the differential entropies of three random
vectors: $\Xv$, $\Xv+\sqrt{a}\Zv$ and $\Xv+\Zv$. To apply, the
increments of $\Xv+\sqrt{a}\Zv$ and $\Xv+\Zv$ over $\Xv$ need to be
Gaussian and have \emph{proportional} covariance matrices. For some
applications in network information theory (as we will see shortly),
the proportionality requirement may turn out to be overly
restrictive. A main contribution of this paper is to prove a natural
generalization of Costa's EPI \eqref{eq:SC-epi} by replacing the
real scalar $a$ with a positive semidefinite \emph{matrix}
parameter. The result is summarized in the following theorem.

\begin{theorem}[Generalized Costa's EPI]
Let $\Zv$ be a Gaussian random $n$-vector with a positive definite
covariance matrix $\Nv$, and let $\Av$ be an $n \times n$ real
symmetric matrix such that $0\preceq\Av\preceq\Iv$. Here, $\Iv$
denotes the $n \times n$ identity matrix, and ``$\preceq$" denotes
``less or equal to" in the positive semidefinite partial ordering
between real symmetric matrices. Then,
\begin{align}
\exp  \left[\frac{2}{n}h(\Xv+\Av^{\frac{1}{2}}\Zv)\right] &\geq
|\Iv-\Av|^{\frac{1}{n}}\exp\left[\frac{2}{n}h(\Xv)\right]+
|\Av|^{\frac{1}{n}}\exp\left[\frac{2}{n}h(\Xv+\Zv)\right]
\label{eq:VC-epi}
\end{align}
for any random $n$-vector $\Xv$ independent of $\Zv$. The equality
holds if $\Zv$ is Gaussian and with a covariance matrix $\Bv$ such
that $\Bv-\Av\Bv$ and $\Bv+\Av^{\frac{1}{2}}\Nv\Av^{\frac{1}{2}}$
are proportional. \label{thm:VC-epi}
\end{theorem}

Note that when $\Av=a\Iv$, the generalized Costa EPI \eqref{eq:VC-epi} reduces
to the original Costa EPI \eqref{eq:SC-epi}. On the other hand, when $\Av$ is
not a scaled identity, the covariance matrices of increments of
$\Xv+\Av^{\frac{1}{2}}\Zv$ and $\Xv+\Zv$ over $\Xv$ do not need to be
proportional. As we will see, the ability to cope with a \emph{general} matrix
parameter makes the generalized Costa EPI more flexible and powerful than the
original Costa EPI.

A different but related generalization of Costa's EPI was considered by
Payar\'{o} and Palomar \cite{PP-IT08}, where they examined the concavity of the
entropy-power $\exp \left[\frac{2}{n}h(\Av^{\frac{1}{2}}\Xv+\Zv)\right]$ with
respect to the matrix parameter $\Av$. This line of research was motivated by
the observation that the original Costa EPI \eqref{eq:SC-epi} is equivalent to
the concavity of the entropy power
$\exp\left[\frac{2}{n}h(\sqrt{a}\Xv+\Zv)\right]$ with respect to the scalar
parameter $a$. Unlike the scalar case, Payar\'{o} and Palomar \cite{PP-IT08}
showed that the entropy-power $\exp
\left[\frac{2}{n}h(\Av^{\frac{1}{2}}\Xv+\Zv)\right]$ is in general \emph{not}
concave with respect to the matrix parameter $\Av$. However, the concavity does
hold when $\Av$ is restricted to be \emph{diagonal} \cite{PP-IT08}.

In information theory, a main application of the EPI is to derive extremal
entropy inequalities, which can then be used to solve network communication
problems. In their work \cite{LV-IT07}, Liu and Viswanath derived an extremal
entropy inequality based on the classical EPI of Shannon \cite{Sha-BSTJ48} and
Stam \cite{Sta-IC59} and used it to establish the private message capacity
region of the vector Gaussian broadcast channel via the Marton outer bound
\cite[Theorem~5]{Mar-IT77}. In this paper, we will derive a new extremal
entropy inequality based on the generalized Costa EPI and use it to
characterize the secrecy capacity regions of the degraded vector Gaussian
broadcast channel with layered confidential messages.

The rest of the paper is organized as follows. In Section~\ref{sec:sum}, we
summarize the main results of the paper, including a new extremal entropy
inequality and its applications on the degraded vector Gaussian broadcast
channel with layered confidential messages. In Section~\ref{sec:VC-epi}, we
prove the generalized Costa EPI, following a perturbation approach via a
fundamental relationship between the derivative of mutual information and MMSE
estimate in linear vector Gaussian channels \cite[Theorem~2]{PV-IT06}. In
Section~\ref{sec:LCVC-epi}, we derive the new extremal entropy inequality from
the generalized Costa EPI. The coding theorems for the degraded vector Gaussian
broadcast channel with layered confidential messages are proved in
Section~\ref{sec:App1} and Section~\ref{sec:App2}. Finally, in
Section~\ref{sec:Con}, we conclude the paper with some remarks.

\section{Summary of Main Results}\label{sec:sum}
The following notation will be used throughout the paper. A random vector is
denoted with an upper-case letter (e.g., $\Xv$), its realization is denoted
with the corresponding lower-case letter (e.g., $\xv$), and its probability
density function is denoted with $p(\xv)=p_{\Xv}(\xv)$. We use $\E[\Xv]$ to
denote the expectation of $\Xv$. Thus, the covariance matrix of $\Xv$ is given
by
\begin{align*}
\Cov(\Xv)=\E\left[(\Xv-\E[\Xv])(\Xv-\E[\Xv])^{\T}\right].
\end{align*}
Given any jointly distributed random vectors $(\Xv,\Yv)$, the MMSE
estimate of $\Xv$ from the observation $\Yv$ is the conditional mean
$\E[\Xv|\Yv]$. The MMSE (matrix) is given by:
\begin{align*}
\Cov(\Xv|\Yv)=\E\left[(\Xv-\E[\Xv|\Yv])(\Xv-\E[\Xv|\Yv])^{\T}\right].
\end{align*}

\subsection{A New Extremal Entropy Inequality}
The following extremal entropy inequality is a consequence of the generalized
Costa EPI.

\begin{theorem} \label{thm:LCVC-epi}
Let $\Zv_k$, $k=0,\ldots,K$, be a total of $K+1$ Gaussian random
$n$-vectors with positive definite covariance matrices $\Nv_k$,
respectively. Assume that $\Nv_1 \preceq \ldots \preceq \Nv_K$. If
there exists an $n \times n$ positive semidefinite matrix $\Bv^*$
such that
\begin{align}
\sum_{k=1}^{K}\mu_k(\Bv^*+\Nv_k)^{-1}+\Mv_1=(\Bv^*+\Nv_0)^{-1}+\Mv_2
\label{eq:1}
\end{align}
for some $n \times n$ positive semidefinite matrices $\Mv_1$,
$\Mv_2$ and $\Sv$ with
\begin{align}
\Bv^*\Mv_1&=0\label{eq:2}\\
\mbox{and} \quad \quad (\Sv-\Bv^*)\Mv_2&=0 \label{eq:3}
\end{align}
and real scalars $\mu_k \ge 0$ with $\sum_{k=1}^{K}\mu_k=1$, then
\begin{align}
\sum_{k=1}^K \mu_k h(\Xv+\Zv_k|U)-h(\Xv+\Zv_0|U)&\leq \sum_{k=1}^K
\frac{\mu_k}{2}\log|\Bv^*+\Nv_k|-\frac{1}{2}\log|\Bv^*+\Nv_0|\label{eq:LCVC-epi}
\end{align}
for any $(\Xv,U)$ independent of $(\Zv_0,\ldots,\Zv_K)$ such that
$\E[\Xv\Xv^\T] \preceq \Sv$.
\end{theorem}

Note that \eqref{eq:1}--\eqref{eq:3} are precisely the Karush-Kuhn-Tucker (KKT)
conditions (see \cite[Appendix~IV]{WSS-IT06} and
\cite[Section~5.2]{BNO-book03}) for the optimization program:
\begin{align*}
\max_{0\preceq\Bv\preceq\Sv}\left[\sum_{k=1}^K
\frac{\mu_k}{2}\log|\Bv+\Nv_k|-\frac{1}{2}\log|\Bv+\Nv_0|\right].
\end{align*}
Therefore, \eqref{eq:LCVC-epi} implies that a jointly
\emph{Gaussian} $(U,\Xv)$ such that for each $U=u$, $\Xv$ has the
\emph{same} covariance matrix is an optimal solution to the
optimization program:
\begin{align*}
\max_{(U,\Xv)}\left[\sum_{k=1}^K \mu_k
h(\Xv+\Zv_k|U)-h(\Xv+\Zv_0|U)\right]
\end{align*}
where the maximization is over all $(U,\Xv)$ independent of
$(\Zv_0,\ldots,\Zv_K)$ such that $\E[\Xv\Xv^\T] \preceq \Sv$. Note
that when $K=1$, this is a special case of \cite[Theorem~8]{LV-IT07}
with $\mu=1$.

\subsection{Applications on the Degraded Vector Gaussian Broadcast
Channel with Layered Confidential Messages}

\begin{figure}[t]
  \begin{minipage}[b]{0.48\linewidth}
  \centerline{\includegraphics[width=\linewidth,draft=false]{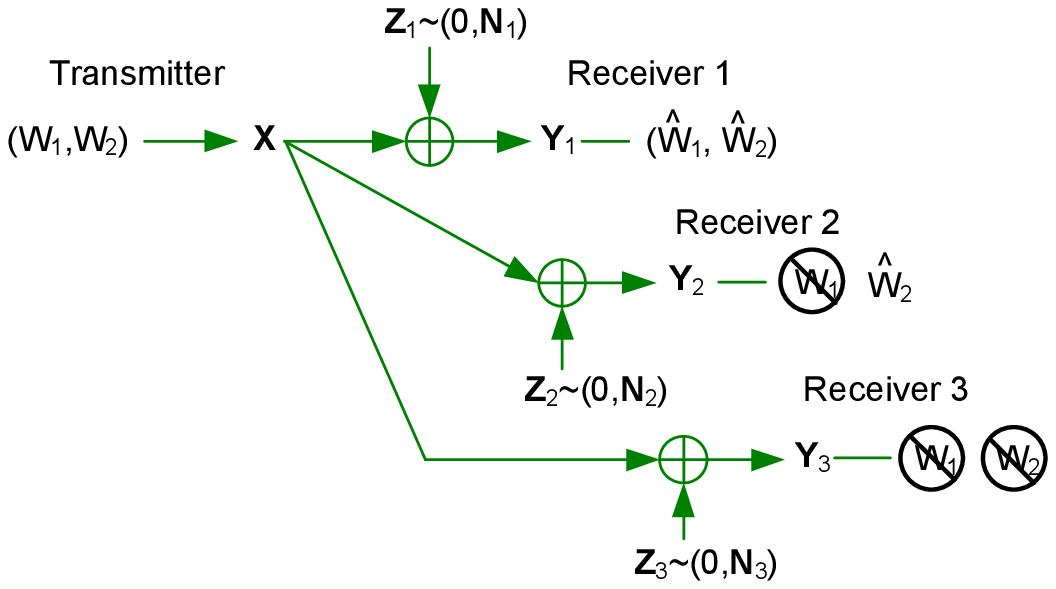}}
    \centerline{\mbox{\small (a) Communication scenario 1}}
      \vspace{0.5cm}
  \end{minipage}
  \begin{minipage}[b]{0.48\linewidth}
  \centerline{\includegraphics[width=\linewidth,draft=false]{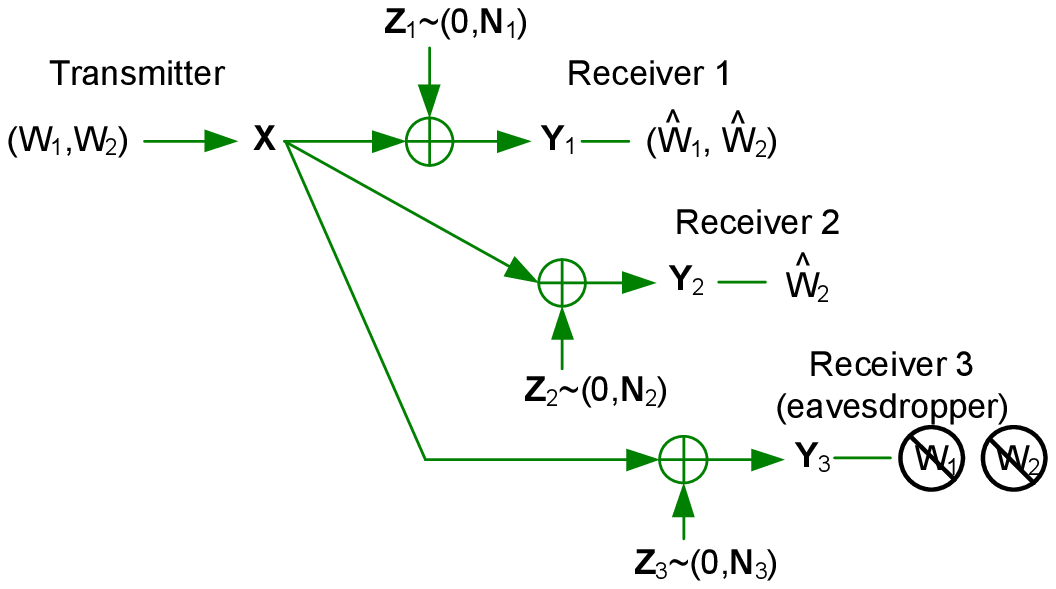}}
    \centerline{\mbox{\small (b) Communication scenario 2}}
              \vspace{0.5cm}
  \end{minipage}
 \caption {Degraded vector Gaussian broadcast channel with layered confidential messages}
  \label{fig:gbc}
\end{figure}

Consider the following vector Gaussian broadcast channel with three receivers:
\begin{align}
\Yv_{k}[t] = \Xv[t]+\Zv_{k}[t], \quad k=1,2,3 \label{eq:Ch}
\end{align}
where $\{\Zv_{k}[t]\}_t$, $k=1,2,3$, are independent and identically
distributed additive vector Gaussian noise processes with zero means
and positive definite covariance matrices $\mathbf{N}_k$,
respectively. The channel input $\{\Xv[t]\}_t$ is subject to a
matrix constraint:
\begin{equation}
\frac{1}{n}\sum_{t=1}^{n}\Xv[t]\Xv^{\T}[t] \preceq \mathbf{S}
\label{eq:cons}
\end{equation}
where $\mathbf{S}$ is a positive semidefinite matrix, and $n$ is the block
length. We assume that the noise covariance matrices are ordered as
\begin{equation}
\Nv_1 \preceq \Nv_2 \preceq \Nv_3, \label{eq:deg}
\end{equation}
i.e., the received signal $\Yv_3[t]$ is (stochastically) degraded
with respect to $\Yv_2[t]$, which is further degraded with respect
to $\Yv_1[t]$.

We consider two different communication scenarios, both with two
independent messages $W_1$ and $W_2$. In the first scenario (see
Fig.~\ref{fig:gbc}-(a)), message $W_1$ is intended for receiver 1
but needs to be kept secret from receivers 2 and 3, and message
$W_2$ is intended for receivers 1 and 2 but needs to be kept
confidential from receiver 3. In the second scenario (see
Fig.~\ref{fig:gbc}-(b)), message $W_1$ is intended for receivers 1
but needs to be kept secret from receiver receiver 3, and message
$W_2$ is intended for receivers 1 but needs to be kept secret from
receiver 3. The confidentiality of the messages at the unintended
receivers is measured using the normalized information-theoretic
criteria \cite{Wyn-BSTJ75,CK-IT78}:
\begin{align}
&\frac{1}{n}I(W_1;\Yv_2^n) \rightarrow 0, \quad
\frac{1}{n}I(W_1;\Yv_3^n) \rightarrow 0, \quad \text{and}  \quad
\frac{1}{n}I(W_2;\Yv_3^n) \rightarrow 0 \label{eq:eqv-r1}
\end{align}
for the first scenario and
\begin{align}
&\frac{1}{n}I(W_1;\Yv_3^n) \rightarrow 0, \quad \text{and}  \quad
\frac{1}{n}I(W_2;\Yv_3^n) \rightarrow 0 \label{eq:eqv-r2}
\end{align}
for the second scenario. Here, the limits are taken as the block
length $n \rightarrow \infty$. The goal is to characterize the
entire secrecy rate region $\mathcal{C}_s=\{(R_1,R_2)\}$ that can be
achieved by any coding scheme.

To characterize the secrecy capacity regions, we will first consider the
discrete memoryless version of the problem with transition probability
$p(y_1,y_2,y_3|x)$ and degradedness order
\begin{equation}
X \rightarrow Y_1 \rightarrow Y_2 \rightarrow Y_3. \label{eq:deg2}
\end{equation}
We have the following single-letter characterizations of the secrecy
capacity regions.

\begin{theorem} \label{thm:dsbc1}
The secrecy capacity region of the discrete memoryless broadcast channel
$p(y_1,y_2,y_3|x)$ with confidential messages $W_1$ (intended for receiver 1
but needs to be kept secret from receivers 2 and 3) and $W_2$ (intended for
receivers 1 and 2 but needs to be kept secret from receiver 3) under the
degradedness order \eqref{eq:deg2} is given by the set of nonnegative rate
pairs $(R_1,R_2)$ such that
\begin{align}
&           & R_1 \leq &\; I(X;Y_1|U)-I(X;Y_2|U) & \notag \\
& \text{and}& R_2 \leq &\; I(U;Y_2)-I(U;Y_3)& \label{eq:rr1}
\end{align}
for some jointly distributed $(U,X)$ satisfying the Markov relation
$$U \rightarrow X \rightarrow (Y_1,Y_2,Y_3).$$
\end{theorem}

\begin{theorem}[{\cite[Theorem~2]{BMK-ITsumb08}}] \label{thm:dsbc2}
The secrecy capacity region of the discrete memoryless broadcast channel
$p(y_1,y_2,y_3|x)$ with confidential messages $W_1$ (intended for receiver 1
but needs to be kept secret from receiver 3) and $W_2$ (intended for receivers
1 and 2 but needs to be kept secret from receiver 3) under the degradedness
order \eqref{eq:deg2} is given by the set of nonnegative rate pairs $(R_1,R_2)$
such that
\begin{align}
&           & R_1 \leq &\; I(X;Y_1|U)-I(X;Y_3|U) & \notag \\
& \text{and}& R_2 \leq &\; I(U;Y_2)-I(U;Y_3)& \label{eq:rr2}
\end{align}
for some jointly distributed $(U,X)$ satisfying the Markov relation
$$U \rightarrow X \rightarrow (Y_1,Y_2,Y_3).$$
\end{theorem}

A proof of Theorem~\ref{thm:dsbc2} can be found in \cite{BMK-ITsumb08}.
Theorem~\ref{thm:dsbc1} can be proved in a similar fashion; for completeness, a
proof is included in Appendix~\ref{app:app}. For the vector Gaussian broadcast
channel \eqref{eq:Ch} under the degradedness order \eqref{eq:deg}, the
single-letter expressions \eqref{eq:rr1} and \eqref{eq:rr2} can be further
evaluated using the extremal entropy inequality \eqref{eq:LCVC-epi}. The
results are summarized in the following theorems.

\begin{theorem} \label{thm:DLMIMO1}
The secrecy capacity region of the vector Gaussian broadcast channel
\eqref{eq:Ch} with confidential messages $W_1$ (intended for
receiver 1 but needs to be kept secret from receivers 2 and 3) and
$W_2$ (intended for receivers 1 and 2 but needs to be kept secret
from receiver 3) and degradedness order \eqref{eq:deg} under the
matrix constraint \eqref{eq:cons} is given by the set of nonnegative
secrecy rate pairs $(R_1,R_2)$ such that
\begin{align}
&           & R_1 & \leq \;
\frac{1}{2}\log\left|\frac{\Bv+\Nv_1}{\Nv_1}\right|-
\frac{1}{2}\log\left|\frac{\Bv+\Nv_2}{\Nv_2}\right|& \notag \\
&\text{and} & R_2 & \leq \;
\frac{1}{2}\log\left|\frac{\Sv+\Nv_2}{\Bv+\Nv_2}\right|-
\frac{1}{2}\log\left|\frac{\Sv+\Nv_3}{\Bv+\Nv_3}\right|&
\label{eq:rr1a}
\end{align}
for some $0 \preceq \Bv \preceq \Sv$.
\end{theorem}

\begin{theorem} \label{thm:DLMIMO2}
The secrecy capacity region of the vector Gaussian broadcast channel
\eqref{eq:Ch} with confidential messages $W_1$ (intended for
receiver 1 but needs to be kept secret from receiver 3) and $W_2$
(intended for receivers 1 and 2 but needs to be kept secret from
receiver 3) and degradedness order \eqref{eq:deg} under the matrix
constraint \eqref{eq:cons} is given by the set of nonnegative
secrecy rate pairs $(R_1,R_2)$ such that
\begin{align}
&           & R_1 & \leq \;
\frac{1}{2}\log\left|\frac{\Bv+\Nv_1}{\Nv_1}\right|-
\frac{1}{2}\log\left|\frac{\Bv+\Nv_3}{\Nv_3}\right|& \notag \\
&\text{and} & R_2 & \leq \;
\frac{1}{2}\log\left|\frac{\Sv+\Nv_2}{\Bv+\Nv_2}\right|-
\frac{1}{2}\log\left|\frac{\Sv+\Nv_3}{\Bv+\Nv_3}\right|&
\label{eq:rr2a}
\end{align}
for some $0 \preceq \Bv \preceq \Sv$.
\end{theorem}

\section{Proof of Theorem~\ref{thm:VC-epi}} \label{sec:VC-epi}
In this section, we prove the generalized Costa EPI \eqref{eq:VC-epi} as stated
in Theorem~\ref{thm:VC-epi}. We first examine the equality condition. Note that
when $\Xv$ is Gaussian, the generalized Costa EPI \eqref{eq:VC-epi} becomes the
matrix inequality:
\begin{align*}
|\Bv+\Av^{\frac{1}{2}}\Nv\Av^{\frac{1}{2}}|^{\frac{1}{n}} &\ge
|\Bv-\Av\Bv|^{\frac{1}{n}}+|\Av\Bv+\Av\Nv|^{\frac{1}{n}}.
\end{align*}
Suppose that $\Bv-\Av\Bv$ and
$\Bv+\Av^{\frac{1}{2}}\Nv\Av^{\frac{1}{2}}$ are proportional, i.e.,
there exists a real scalar $c$ such that
\begin{align*}
\Bv+\Av^{\frac{1}{2}}\Nv\Av^{\frac{1}{2}}=c(\Bv-\Av\Bv).
\end{align*}
Since both matrices $\Av$ and $\Bv$ are symmetric, this implies that
$\Av\Bv$ is also symmetric, i.e.,
$$\Av\Bv=\Bv^{\T}\Av^{\T}=\Bv\Av.$$ Therefore, $\Av$ and $\Bv$ must
have the \emph{same} eigenvector matrix \cite{Gilbert} and hence
\begin{align*}
\Av\Bv &=\Av^{\frac{1}{2}}\Bv\Av^{\frac{1}{2}}.
\end{align*}
It follows that
\begin{align*}
\Av^{\frac{1}{2}}\Bv\Av^{\frac{1}{2}}+\Av^{\frac{1}{2}}\Nv\Av^{\frac{1}{2}}&=
\Bv+\Av^{\frac{1}{2}}\Nv\Av^{\frac{1}{2}}-(\Bv-\Av\Bv)\\
&=(c-1)(\Bv-\Av\Bv)
\end{align*}
i.e.,
$\Av^{\frac{1}{2}}\Bv\Av^{\frac{1}{2}}+\Av^{\frac{1}{2}}\Nv\Av^{\frac{1}{2}}$
and $\Bv-\Av\Bv$ are proportional. Therefore,
\begin{align*}
|\Bv+\Av^{\frac{1}{2}}\Nv\Av^{\frac{1}{2}}|^{\frac{1}{n}}&=
|\Bv-\Av\Bv+(\Av^{\frac{1}{2}}\Bv\Av^{\frac{1}{2}}+\Av^{\frac{1}{2}}\Nv\Av^{\frac{1}{2}})|^{\frac{1}{n}}\\
&=|\Bv-\Av\Bv|^{\frac{1}{n}}+|\Av^{\frac{1}{2}}\Bv\Av^{\frac{1}{2}}+\Av^{\frac{1}{2}}\Nv\Av^{\frac{1}{2}}|^{\frac{1}{n}}\\
&=|\Bv-\Av\Bv|^{\frac{1}{n}}+|\Av\Bv+\Av\Nv|^{\frac{1}{n}}.
\end{align*}
This proved the desired equality condition.

We now turn to the proof of the inequality. First consider the
special case when $|\Av|=0$. Since
\begin{align*}
h(\Xv+\Av^{\frac{1}{2}}\Zv)-h(\Xv)=I(\Av^{\frac{1}{2}}\Zv;\Xv+\Av^{\frac{1}{2}}\Zv)
\geq 0,
\end{align*}
we have
\begin{align*}
\exp\left[\frac{2}{n}h(\Xv+\Av^{\frac{1}{2}}\Zv)\right] &\geq
\exp\left[\frac{2}{n}h(\Xv)\right]\\
&\geq |\Iv-\Av|^{\frac{1}{n}}\exp\left[\frac{2}{n}h(\Xv)\right]
\end{align*}
where the last inequality follows from the assumption that
$0\preceq\Av\preceq\Iv$ and hence $0 \leq |\Iv-\Av| \leq 1$.

Next, consider the general case when $|\Av|>0$. The proof is rather
long so we divide it into several steps.

\emph{Step 1--Constructing a monotone path.} To prove the
generalized Costa EPI \eqref{eq:VC-epi}, we can equivalently show
that
\begin{align}
\exp\left[\frac{2}{n}h(\Xv+\Zv)\right] &\leq
|\Av|^{-\frac{1}{n}}\exp\left[\frac{2}{n}h(\Xv+\Av^{\frac{1}{2}}\Zv)\right]-\left(
\frac{|\Iv-\Av|}{|\Av|}\right)^{\frac{1}{n}}\exp\left[\frac{2}{n}h(\Xv)\right].
\label{eq:App-ep}
\end{align}
Since $\Xv$ and $\Zv$ are independent, we have
\begin{align}
h( \Xv +\Av^{\frac{1}{2}}\Zv)-h(\Xv)&=
h(\Av^{-\frac{1}{2}}\Xv+\Zv)-h(\Av^{-\frac{1}{2}}\Xv)\notag\\
&=h(\Av^{-\frac{1}{2}}\Xv+\Zv)-h(\Av^{-\frac{1}{2}}\Xv|\Zv)\notag\\
&= I(\Zv;\Av^{-\frac{1}{2}}\Xv+\Zv)\label{eq:I1}
\end{align}
and
\begin{equation}
h(\Xv+\Zv)-h(\Xv) =I(\Zv;\Xv+\Zv).\label{eq:I2}
\end{equation}
Divide both sides of \eqref{eq:App-ep} by $\exp\left[\frac{2}{n}h(\Xv)\right]$
and use \eqref{eq:I1} and \eqref{eq:I2}. Then, \eqref{eq:App-ep} can be
equivalently written as
\begin{align}
\exp  \left[\frac{2}{n}I(\Zv;\Xv+\Zv)\right] &\leq
|\Av|^{-\frac{1}{n}}\left\{\exp\left[\frac{2}{n}I(\Zv;
\Av^{-\frac{1}{2}}\Xv+\Zv)\right]-|\Iv-\Av|^{\frac{1}{n}}\right\}.
\label{eq:D-epi}
\end{align}
Let
\begin{align}
F(\Dv)&:= |\Dv|^{\frac{2}{n}}
\Biggl\{\exp\left[\frac{2}{n}I(\Zv;\Dv\Xv+\Zv)\right]-|\Iv-\Dv^{-2}|^{\frac{1}{n}}\Biggr\}.
\label{eq:def-FF}
\end{align}
With this definition, \eqref{eq:D-epi} can be equivalently written as
\begin{align}
F(\Iv) \leq F(\Av^{-\frac{1}{2}}). \label{eq:Der0}
\end{align}

To show the inequality \eqref{eq:Der0}, it is sufficient to construct a family
of $n \times n$ positive definite matrices $\{\Dv(\gamma)\}_{\gamma}$
connecting $\Iv$ and $\Av^{-\frac{1}{2}}$ such that $F(\Dv(\gamma))$ is
monotone along the path. Unlike the scalar case where there is only one path
connecting $1$ to $1/\sqrt{a}$, in the matrix case there are infinitely many
paths connecting $\Iv$ and $\Av^{-\frac{1}{2}}$. Here, we consider the special
choice
\begin{equation}
\Dv(\gamma) := \left[\Iv+\gamma
(\Av^{-1}-\Iv)\right]^{\frac{1}{2}}\label{eq:def-Dg}
\end{equation}
and show that
\begin{equation}
\frac{\partial F}{\partial\gamma} \geq 0, \quad \forall \gamma \in
[0,1]. \label{eq:Der}
\end{equation}
along this particular path.

\emph{Step 2--Calculating the derivative $\frac{\partial
F}{\partial\gamma}$.} Following \cite[Theorem~5]{GSV-ISIT06}, we
have
\begin{equation*}
I(\Zv; \Dv \Xv +\Zv)=I(\Xv;\Dv \Xv +\Zv)+h(\Zv)- h(\Xv)-\log|\Dv|
\end{equation*}
and
\begin{equation*}
\Cov(\Xv|\Dv \Xv +\Zv)=\Dv^{-1}\, \Cov(\Zv|\Dv \Xv +\Zv)
\Dv^{-\T}.\label{eq:C-C}
\end{equation*}
Let $\Nv:=\Cov(\Zv)$ and note that $\Dv$ is symmetric. We have
\begin{align}
\frac{\partial}{\partial \Dv}I(\Zv; \Dv\Xv+\Zv)&  =
\frac{\partial}{\partial \Dv}I(\Xv; \Dv\Xv+\Zv)-\Dv^{-1}\nonumber\\
& = \Nv^{-1} \Dv\, \Cov(\Xv|\Dv \Xv
+\Zv)-\Dv^{-1}\notag\\
&=\left(\Nv^{-1} \Cov(\Zv|\Dv \Xv
+\Zv)-\Iv\right)\Dv^{-1}\label{eq:mB0}
\end{align}
where the second equality follows from the fundamental relationship
between the derivative of mutual information and MMSE estimate in
linear vector Gaussian channels as stated in
\cite[Theorem~2]{PV-IT06}.

From \eqref{eq:mB0}, the derivative $\frac{\partial F}{\partial\Dv}$ can be
calculated as
\begin{align}
\frac{\partial F}{\partial\Dv}
=&\frac{2}{n}|\Dv|^{\frac{2}{n}}\Dv^{-1}\Biggl\{\exp\left[\frac{2}{n}I(\Zv;\Dv\Xv+\Zv)\right]-
|\Iv-\Dv^{-2}|^{\frac{1}{n}}\Bigg\}+\notag\\
&|\Dv|^{\frac{2}{n}}\Biggl\{\frac{2}{n}\exp\left[\frac{2}{n}I(\Zv;\Dv\Xv+\Zv)\right]\frac{\partial
I(\Zv; \Dv \Xv +\Zv)}{\partial \Dv}-\frac{2}{n}|\Iv-\Dv^{-2}|^{\frac{1}{n}}(\Iv-\Dv^{-2})^{-1} \Dv^{-3}\Biggr\}\notag\\
=&\frac{2}{n}|\Dv|^{\frac{2}{n}}\Biggl\{\left\{\exp\left[\frac{2}{n}I(\Zv;\Dv\Xv+\Zv)\right]-
|\Iv-\Dv^{-2}|^{\frac{1}{n}}\right\}\Iv+\notag\\
&\exp\left[\frac{2}{n}I(\Zv;\Dv\Xv+\Zv)\right](\Nv^{-1} \Cov(\Zv|\Dv
\Xv +\Zv)-\Iv )-|\Iv-\Dv^{-2}|^{\frac{1}{n}}(\Dv^{2}-\Iv)^{-1}\Biggr\}\Dv^{-1}\notag\\
=&\frac{2}{n}|\Dv|^{\frac{2}{n}}\Biggl\{\exp\left[\frac{2}{n}I(\Zv;\Dv\Xv+\Zv)\right]\Nv^{-1}
\Cov(\Zv|\Dv \Xv +\Zv)-
|\Iv-\Dv^{-2}|^{\frac{1}{n}}\left[\Iv+(\Dv^{2}-\Iv)^{-1}\right]\Biggr\}\Dv^{-1}.
\label{eq:dFD}
\end{align}
The derivative $\frac{\partial\Dv}{\partial\gamma}$ can be calculated as
\begin{align}
\frac{\partial \Dv}{\partial\gamma}&=\frac{1}{2}\left[\Iv+\gamma
(\Av^{-1}-\Iv)\right]^{-\frac{1}{2}}(\Av^{-1}-\Iv)\notag\\
&=\frac{1}{2\gamma}\Dv^{-1}(\Dv^2-\Iv)\notag \\
&=\frac{1}{2\gamma}\Dv(\Iv-\Dv^{-2}). \label{eq:dDdg}
\end{align}
By \eqref{eq:dFD}, \eqref{eq:dDdg} and the chain rule of differentiation
\cite[Chapter~17.5]{Seb-book08},
\begin{align}
\frac{\partial F}{\partial\gamma}&=\tr \left\{\frac{\partial
F}{\partial\Dv}\, \frac{\partial \Dv}{\partial
\gamma}\right\}\notag\\
&=\frac{|\Dv|^{\frac{2}{n}}}{n}\tr\Biggl\{\left[\exp\left[\frac{2}{n}I(\Zv;\Dv\Xv+\Zv)\right]\Nv^{-1}
\Cov(\Zv|\Dv \Xv +\Zv)- |\Iv-\Dv^{-2}|^{\frac{1}{n}}
\left[\Iv+(\Dv^{2}-\Iv)^{-1}\right]\right]\frac{\Iv-\Dv^{-2}}{\gamma}\Biggr\}\notag \\
&=\frac{|\Dv|^{\frac{2}{n}}}{n\gamma}\tr\left\{\exp\left[\frac{2}{n}I(\Zv;\Dv\Xv+\Zv)\right]\Nv^{-1}
\Cov(\Zv|\Dv \Xv +\Zv)(\Iv-\Dv^{-2})- |\Iv-\Dv^{-2}|^{\frac{1}{n}} \Iv\right\} \notag \\
&=\frac{|\Dv|^{\frac{2}{n}}}{n\gamma}\left\{\exp\left[\frac{2}{n}I(\Zv;\Dv\Xv+\Zv)\right]\tr\left\{\Nv^{-1}
\Cov(\Zv|\Dv \Xv +\Zv)(\Iv-\Dv^{-2})\right\}-n
|\Iv-\Dv^{-2}|^{\frac{1}{n}}\right\}.\label{eq:dfdg}
\end{align}

\emph{Step 3--Proving $\frac{\partial F}{\partial\gamma} \geq 0$.}
The mutual information $I(\Zv;\Dv\Xv+\Zv)$ can be bounded from below
as follows:
\begin{align}
I(\Zv;\Dv\Xv+\Zv)&\geq  I(\Zv; \E[\Zv|\Dv\Xv+\Zv])\notag\\
&  =  h(\Zv)-h(\Zv|\E[\Zv|\Dv\Xv+\Zv])\nonumber\\
& =  \frac{1}{2}\log(2\pi e)^n |\Nv|-h(\Zv-\E[\Zv|\Dv\Xv+\Zv]|\E[\Zv|\Dv\Xv+\Zv])\nonumber\\
&  \geq  \frac{1}{2}\log(2\pi e)^n |\Nv|-h(\Zv-\E[\Zv|\Dv\Xv+\Zv])\nonumber\\
&  \geq  \frac{1}{2}\log(2\pi e)^n |\Nv|- \frac{1}{2}\log(2\pi
e)^n \bigl|\Cov(\Zv|\Dv \Xv +\Zv)\bigr|\notag\\
&  = \frac{1}{2}\log \frac{|\Nv|}{|\Cov(\Zv|\Dv \Xv
+\Zv)|}.\label{eq:mC}
\end{align}
Here, the first inequality follows from the Markov relation
$$\Zv \rightarrow \Dv\Xv+\Zv \rightarrow \E[\Zv|\Dv\Xv+\Zv]$$ and
the chain rule of mutual information \cite[Chapter~2.8]{Cover}; the second
inequality follows from the fact that conditioning reduces differential entropy
\cite[Chapter~9.6]{Cover}; and the third inequality follows from the well-known
fact that Gaussian maximizes differential entropy for a given covariance matrix
\cite[Chapter~9.6]{Cover}. By \eqref{eq:mC},
\begin{align}
|\Iv-\Dv^{-2}|^{\frac{1}{n}}\exp\left[-\frac{2}{n}I(\Zv;\Dv\Xv+\Zv)\right]
& \leq \; |\Nv^{-1}\Cov(\Zv|\Dv \Xv
+\Zv)(\Iv-\Dv^{-2})|^{\frac{1}{n}}\nonumber\\
&\leq \; \frac{1}{n} \tr\left\{\Nv^{-1} \Cov(\Zv|\Dv \Xv
+\Zv)(\Iv-\Dv^{-2})\right\} \label{eq:B-m1}
\end{align}
where the last inequality follows from the well-known inequality of
arithmetic and geometric means \cite[p.~136]{GS-book93}.

Finally, substituting \eqref{eq:B-m1} into \eqref{eq:dfdg} establishes the fact
that $\frac{\partial F}{\partial\gamma} \geq 0$ for all $\gamma \in [0,1]$. In
particular, we have $F(\Dv(1)) \geq F(\Dv(0))$. This proved the desired
inequality \eqref{eq:D-epi} and hence the generalized Costa EPI
\eqref{eq:VC-epi}.

\section{Proof of Theorem~\ref{thm:LCVC-epi}} \label{sec:LCVC-epi}
In this section, we prove the extremal entropy inequality
\eqref{eq:LCVC-epi} as stated in Theorem~\ref{thm:LCVC-epi}. We will
first state a series of corollaries of Theorem~\ref{thm:VC-epi}, as
intermediate results leading to Theorem~\ref{thm:LCVC-epi}. Based on
the final corollary, we will prove Theorem~\ref{thm:LCVC-epi} using
an \emph{enhancement} argument.

\begin{corollary} \label{cor:cond}
Let $\Zv$ be a Gaussian random $n$-vector with a positive definite
covariance matrix, and let $\Av$ be an $n\times n$ positive real
symmetric matrix such that $0\preceq\Av\preceq \Iv$. Then
\begin{align}
\exp\left[\frac{2}{n}h(\Xv+\Av^{\frac{1}{2}}\Zv|U)\right]&\geq
|\Iv-\Av|^{\frac{1}{n}}\exp\left[\frac{2}{n}h(\Xv|U)\right]+
|\Av|^{\frac{1}{n}}\exp\left[\frac{2}{n}h(\Xv+\Zv|U)\right]
\label{eq:CVC-epi}
\end{align}
for any $(\Xv,U)$ independent of $\Zv$.
\end{corollary}

\begin{corollary}  \label{cor:Lepi-2}
Let $\Zv_1$, $\Zv_2$ and $\Zv_3$ be Gaussian random $n$-vectors with
positive definite covariance matrices $\Nv_1$, $\Nv_2$ and $\Nv_3$,
respectively. Assume that $\Nv_1 \preceq \Nv_3$. If there exists an
$n \times n$  positive semidefinite matrix $\Bv^*$ such that
\begin{align}
(\Bv^*+\Nv_1)^{-1}+\mu(\Bv^*+\Nv_3)^{-1}=(1+\mu)(\Bv^*+\Nv_2)^{-1}
\label{eq:LVC-2}
\end{align}
for some real scalar $\mu \ge 0$, then
\begin{align}
h(\Xv+\Zv_1|U)+\mu h&(\Xv+\Zv_3|U)-(1+\mu)h(\Xv+\Zv_2|U)\notag\\
&\leq
\frac{1}{2}\log|\Bv^*+\Nv_1|+\frac{\mu}{2}\log|\Bv^*+\Nv_3|-\frac{1+\mu}{2}\log|\Bv^*+\Nv_2|
\label{eq:Lepi-2}
\end{align}
for any $(\Xv,U)$ independent of $(\Zv_1,\Zv_2,\Zv_3)$.
\end{corollary}

\begin{corollary}  \label{cor:Lepi-K}
Let $\Zv_k$, $k=0,\ldots, K$, be a collection of $K+1$ Gaussian random
$n$-vectors with respective positive definite covariance matrices $\Nv_k$.
Assume that $\Nv_1 \preceq \ldots \preceq \Nv_K$. If there exists an $n \times
n$ positive semidefinite matrix $\Bv^*$ such that
\begin{align}
\sum_{k=1}^{K}\mu_k(\Bv^*+\Nv_k)^{-1}=(\Bv^*+\Nv_0)^{-1}
\label{eq:LVC-K}
\end{align}
for some $\mu_k \ge 0$ with $\sum_{k=1}^{K}\mu_k=1$, then
\begin{align}
\sum_{k=1}^K \mu_k h(\Xv+\Zv_k|U)-h(\Xv+\Zv_0|U)&\leq \sum_{k=1}^K
\frac{\mu_k}{2}\log|\Bv^*+\Nv_k|-\frac{1}{2}\log|\Bv^*+\Nv_0|\label{eq:Lepi-K}
\end{align}
for any $(\Xv,U)$ independent of $(\Zv_0,\ldots,\Zv_K)$.
\end{corollary}

A proof of Corollaries~\ref{cor:cond}, \ref{cor:Lepi-2} and
\ref{cor:Lepi-K} can be found in Appendices~\ref{app:cond},
\ref{app:Lepi-2} and \ref{app:Lepi-K}, respectively. We are now
ready to prove Theorem~\ref{thm:LCVC-epi}. Note that the special
case with $\Mv_1=\Mv_2=0$ was proved in Corollary~\ref{cor:Lepi-K}.
To extend the result of Corollary~\ref{cor:Lepi-K} to nonzero
$\Mv_1$ and $\Mv_2$, we will consider an enhancement argument, which
was first introduced by Weingarten, Steinberg and Shamai in
\cite{WSS-IT06}.

Let $\Nt_1$ and $\Nt_0$ be $n \times n$ real symmetric matrices such
that:
\begin{align}
\mu_1(\Bv^*+\Nt_1)^{-1}&=\mu_1(\Bv^*+\Nv_1)^{-1}+\Mv_1\label{eq:4}\\
\mbox{and} \quad\quad
(\Bv^*+\Nt_0)^{-1}&=(\Bv^*+\Nv_0)^{-1}+\Mv_2.\label{eq:5}
\end{align}
As shown in \cite[Lemma~11~and~12]{WSS-IT06}, $\Nt_1$ and $\Nt_0$
satisfy the following properties:
\begin{align}
0 \prec \Nt_1 &= \left(\Nv_1^{-1}+\mu_1^{-1}\Mv_1\right)^{-1}
\preceq \Nv_1, \label{eq:enh1}
\end{align}
\begin{align}
\Nt_1 &\preceq \Nt_0 \preceq \Nv_0, \label{eq:enh2}
\end{align}
\begin{align}
\left|\frac{\Bv^*+\Nt_1}{\Nt_1}\right|&=\left|\frac{\Bv^*+\Nv_1}{\Nv_1}\right| \label{eq:enh3}
\end{align}
and \begin{align}
\left|\frac{\Sv+\Nt_0}{\Bv^*+\Nt_0}\right|&=\left|\frac{\Sv+\Nv_2}{\Bv^*+\Nv_2}\right|.
\label{eq:enh4}
\end{align}
Let $\Zt_0$ and $\Zt_1$ be two Gaussian $n$-vectors with covariance
matrices $\Nt_0$ and $\Nt_1$, respectively. Note from
\eqref{eq:enh1} that $\Nt_1 \preceq \Nv_1 \preceq \Nv_2 \preceq
\ldots \preceq \Nv_K$. Moreover, substitute \eqref{eq:4} and
\eqref{eq:5} into \eqref{eq:1} and we have
\begin{align}
\mu_1(\Bv^*+\Nt_1)^{-1}+\sum_{k=2}^K\mu_k(\Bv^*+\Nv_k)^{-1}&=(\Bv^*+\Nt_0)^{-1}.
\label{eq:kkt4}
\end{align}
Thus, by Corollary~\ref{cor:Lepi-K}
\begin{align}
\mu_1 h(\Xv+\Zt_1|U)+&\sum_{k=2}^K \mu_k
h(\Xv+\Zv_k|U)-h(\Xv+\Zt_0|U)\notag\\
&\leq \frac{\mu_1}{2}(\Bv^*+\Nt_1)^{-1}+\sum_{k=2}^K
\frac{\mu_k}{2}\log|\Bv^*+\Nv_k|-\frac{1}{2}\log|\Bv^*+\Nt_0|
\label{eq:t1}
\end{align}
for any $(\Xv,U)$ independent of $(\Zt_0,\Zt_1,\Zv_2,\ldots,\Zv_K)$.

On the other hand, note from \eqref{eq:enh1} that $\Nt_1 \preceq
\Nv_1$. We have
\begin{align*}
I(\Xv;\Xv+\Zv_1|U) \le I(\Xv;\Xv+\Zt_1|U)
\end{align*}
for any $(\Xv,U)$ independent of $(\Zv_1,\Zt_1)$. Thus,
\begin{align}
h(\Xv+\Zt_1|U)-h(\Xv+\Zv_1|U)&\ge h(\Zt_1)-h(\Zv_1)\notag\\
&=\frac{1}{2}\log\left|\frac{\Nt_1}{\Nv_1}\right|\notag\\
&=\frac{1}{2}\log\left|\frac{\Bv^*+\Nt_1}{\Bv^*+\Nv_1}\right|
\label{eq:t2}
\end{align}
where the last equality follows from \eqref{eq:enh3}.

Also note from \eqref{eq:enh2} that $\Nt_0 \preceq \Nv_0$. Let
$\hat{\Zv}_0$ be a Gaussian $n$-vector with covariance matrix
$\Nv_0-\Nt_0$ and independent of $(\Zt_0,\Xv,U)$. We have
\begin{align}
h(\Xv+\Zv_0|U)-h(\Xv+\Zt_0|U) &=
h(\Xv+\Zt_0+\hat{\Zv}_0|U)-h(\Xv+\Zt_0|U)\notag\\
&=I(\hat{\Zv}_0;\Xv+\Zt_0+\hat{\Zv}_0|U)\notag\\
&\ge I(\hat{\Zv}_0;\Xv+\Zt_0+\hat{\Zv}_0)\notag\\
&\ge \frac{1}{2}\log\left|\frac{\Cov(\Xv)+\Nv_0}{\Cov(\Xv)+\Nt_0}\right|\notag\\
&\ge
\frac{1}{2}\log\left|\frac{\Sv+\Nv_0}{\Sv+\Nt_0}\right|\label{eq:t3}\\
&=
\frac{1}{2}\log\left|\frac{\Bv^*+\Nv_0}{\Bv^*+\Nt_0}\right|\label{eq:t4}
\end{align}
for any $(\Xv,U)$ independent of $(\Zv_0,\Zt_0)$ such that $\E[\Xv\Xv^\T]
\preceq \Sv$. Here, the first inequality follows from the independence of
$\hat{\Zv}_0$ and $U$; the second inequality follows from the worst noise
result \cite[Lemma~II.2]{DC-IT01}; the third inequality follows from the fact
that $\Nt_0 \preceq \Nv_0$ and $\Cov(\Xv) \preceq \E[\Xv\Xv^\T] \preceq \Sv$;
and the last inequality follows from \eqref{eq:enh4}.

Finally, put together \eqref{eq:t1}, \eqref{eq:t2} and \eqref{eq:t4}
and we may obtain
\begin{align*}
\sum_{k=1}^K \mu_k &h(\Xv+\Zv_k|U)-h(\Xv+\Zv_0|U)\\
&=\left[\mu_1 h(\Xv+\Zt_1|U)+\sum_{k=2}^K \mu_k
h(\Xv+\Zv_k|U)-h(\Xv+\Zt_0|U)\right]-\notag\\
&\hspace{16pt}\mu_1\left[h(\Xv+\Zt_1|U)-h(\Xv+\Zv_1|U)\right]-
\left[h(\Xv+\Zv_0|U)-h(\Xv+\Zt_0|U)\right]\notag\\
&\le \left[\frac{\mu_1}{2}(\Bv^*+\Nt_1)^{-1}+\sum_{k=2}^K
\frac{\mu_k}{2}\log|\Bv^*+\Nv_k|-\frac{1}{2}\log|\Bv^*+\Nt_0|\right]-\notag\\
&\hspace{16pt}
\frac{\mu_1}{2}\log\left|\frac{\Bv^*+\Nt_1}{\Bv^*+\Nv_1}\right|
-\frac{1}{2}\log\left|\frac{\Bv^*+\Nv_0}{\Bv^*+\Nt_0}\right|\notag\\
&= \sum_{k=1}^K
\frac{\mu_k}{2}\log|\Bv^*+\Nv_k|-\frac{1}{2}\log|\Bv^*+\Nv_0|
\end{align*}
for any $(\Xv,U)$ independent of $(\Zv_0,\Zv_1,\ldots,\Zv_K)$ such
that $\E[\Xv\Xv^\T] \preceq \Sv$. This completes the proof of
Theorem~\ref{thm:LCVC-epi}.

\section{Proof of Theorem~\ref{thm:DLMIMO1}} \label{sec:App1}
In this section, we prove Theorem~\ref{thm:DLMIMO1}. Note that the
achievability of the secrecy rate region \eqref{eq:rr1a} can be
obtained from the secrecy rate region \eqref{eq:rr1} by letting
$\Uv$ and $\Vv$ be two independent Gaussian vectors with zero means
and covariance matrices $\Sv-\Bv$ and $\Bv$, respectively and
$\Xv=\Uv+\Vv$. We therefore concentrate on the converse part of the
theorem.

To show that \eqref{eq:rr1a} is indeed the secrecy capacity region
of the vector Gaussian broadcast channel \eqref{eq:Ch}, we will
consider proof by contradiction. Assume that $(R_1^o,R_2^o)$ is an
achievable secrecy rate pair that lies \emph{outside} the secrecy
rate region \eqref{eq:rr1a}. Note that $\Nv_1\preceq\Nv_2$. From
\cite[Theorem~1]{LS-ITSubm08}, we can bound $R_1^o$ by
\begin{align*}
R_1^o \le \frac{1}{2}\log\left|\frac{\Sv+\Nv_1}{\Nv_1}\right|-
\frac{1}{2}\log\left|\frac{\Sv+\Nv_2}{\Nv_2}\right| = R_1^{max}.
\end{align*}
Note that when $R_2^o=0$, $R_1^{max}$ is achievable by letting
$\Bv=\Sv$ in \eqref{eq:rr1}. Thus, we may assume that $R_2^o>0$ and
write $R_1^o=R_1^*+\delta$ for some $\delta>0$ where $R_1^*$ is
given by
\begin{align*}
\max_{\Bv} & \quad \left[\frac{1}{2}\log\left|\frac{\Bv+\Nv_1}{\Nv_1}\right|-
\frac{1}{2}\log\left|\frac{\Bv+\Nv_2}{\Nv_2}\right|\right]\\
\mbox{subject to:}& \quad 0 \preceq \Bv \preceq \Sv\\
& \quad \frac{1}{2}\log\left|\frac{\Sv+\Nv_2}{\Bv+\Nv_2}\right|-
\frac{1}{2}\log\left|\frac{\Sv+\Nv_3}{\Bv+\Nv_3}\right| \ge R_2^o.
\end{align*}
Let $\Bv^*$ be an optimal solution to the above optimization program. Then,
$\Bv^*$ must satisfy the following KKT conditions\footnote{As this optimization
program is not convex, a set of constraint qualifications (CQs) should be
checked to make sure that the KKT conditions indeed hold. The CQs stated in
Appendix~IV of \cite{WSS-IT06} hold in a trivial manner for this program.}:
\begin{align}
(\Bv^*+\Nv_1)^{-1}+\mu(\Bv^*+\Nv_3)^{-1}+\Mv_1&=(1+\mu)(\Bv^*+\Nv_2)^{-1}+\Mv_2\label{eq:kkt1}\\
\Bv^*\Mv_1 & = 0\label{eq:kkt2}\\
\mbox{and} \quad\quad (\Sv-\Bv^*)\Mv_2 & = 0\label{eq:kkt3}
\end{align}
where $\Mv_1$ and $\Mv_2$ are $n \times n$ positive semidefinite
matrices, and $\mu$ is a nonnegative real scalar such that $\mu>0$
if and only if
$$\frac{1}{2}\log\left|\frac{\Sv+\Nv_2}{\Bv^*+\Nv_2}\right|-
\frac{1}{2}\log\left|\frac{\Sv+\Nv_3}{\Bv^*+\Nv_3}\right| = R_2^o.$$
Thus,
\begin{align}
R_1^o+\mu R_2^o =
\left[\frac{1}{2}\log\left|\frac{\Bv^*+\Nv_1}{\Nv_1}\right|-
\frac{1}{2}\log\left|\frac{\Bv^*+\Nv_2}{\Nv_2}\right|\right]+
\mu\left[\frac{1}{2}\log\left|\frac{\Sv+\Nv_2}{\Bv^*+\Nv_2}\right|-
\frac{1}{2}\log\left|\frac{\Sv+\Nv_3}{\Bv^*+\Nv_3}\right|\right]+\delta.
\label{eq:c1}
\end{align}

On the other hand, by the converse part of Theorem~\ref{thm:dsbc1}
\begin{align}
R_1^o+\mu R_2^o \le\;&
[I(\Xv;\Xv+\Zv_1|U)-I(\Xv;\Xv+\Zv_2|U)]+\mu[I(U;\Xv+\Zv_2)-I(U;\Xv+\Zv_3)]\notag\\
=\;&[h(\Zv_2)-h(\Zv_1)]-\mu[h(\Xv+\Zv_3)-h(\Xv+\Zv_2)]+\notag\\
\;&[h(\Xv+\Zv_1|U)+\mu h(\Xv+\Zv_3|U)-(1+\mu)h(\Xv+\Zv_2|U)]\notag\\
=\;&\frac{1}{2}\log\left|\frac{\Nv_2}{\Nv_1}\right|-\mu[h(\Xv+\Zv_3)-h(\Xv+\Zv_2)]+\notag\\
\;&[h(\Xv+\Zv_1|U)+\mu h(\Xv+\Zv_3|U)-(1+\mu)h(\Xv+\Zv_2|U)]
\label{eq:t5}
\end{align}
for some jointly distributed $(U,\Xv)$ independent of
$(\Zv_1,\Zv_2,\Zv_3)$. Note that $\Nv_2 \preceq \Nv_3$. Similar to
\eqref{eq:t3}, we may obtain
\begin{align}
h(\Xv+\Zv_3)-h(\Xv+\Zv_2) &\ge
\frac{1}{2}\log\left|\frac{\Sv+\Nv_3}{\Sv+\Nv_2}\right|.
\label{eq:t6}
\end{align}
Moreover, by letting
\begin{align*}
\mu_1=\frac{1}{1+\mu}, \quad \mu_3=\frac{\mu}{1+\mu}, \quad
\tilde{\Mv}_1=\frac{\Mv_1}{1+\mu}, \quad \mbox{and} \;\;
\tilde{\Mv}_2=\frac{\Mv_2}{1+\mu}
\end{align*}
we can rewrite the KKT conditions \eqref{eq:kkt1}--\eqref{eq:kkt3}
as
\begin{align*}
\mu_1(\Bv^*+\Nv_1)^{-1}+\mu_3(\Bv^*+\Nv_3)^{-1}+\tilde{\Mv}_1&=(\Bv^*+\Nv_2)^{-1}+\tilde{\Mv}_2\\
\Bv^*\tilde{\Mv}_1 & = 0\\
\mbox{and} \quad\quad (\Sv-\Bv^*)\tilde{\Mv}_2 & = 0.
\end{align*}
Thus, by Theorem~\ref{thm:LCVC-epi}
\begin{align}
h(\Xv+\Zv_1|U)+\mu h&(\Xv+\Zv_3|U)-(1+\mu)h(\Xv+\Zv_2|U)\notag\\
&\leq
\frac{1}{2}\log|\Bv^*+\Nv_1|+\frac{\mu}{2}\log|\Bv^*+\Nv_3|-\frac{1+\mu}{2}\log|\Bv^*+\Nv_2|.
\label{eq:t7}
\end{align}
Substituting \eqref{eq:t6} and \eqref{eq:t7} into \eqref{eq:t5}, we
have
\begin{align}
R_1^o+\mu R_2^o \le&\;
\frac{1}{2}\log\left|\frac{\Nv_2}{\Nv_1}\right|-\frac{\mu}{2}\log\left|\frac{\Sv+\Nv_3}{\Sv+\Nv_2}\right|+\notag\\
&\;\left[\frac{1}{2}\log|\Bv^*+\Nv_1|+\frac{\mu}{2}\log|\Bv^*+\Nv_3|-\frac{1+\mu}{2}\log|\Bv^*+\Nv_2|\right]\notag\\
=&\;\left[\frac{1}{2}\log\left|\frac{\Bv^*+\Nv_1}{\Nv_1}\right|-
\frac{1}{2}\log\left|\frac{\Bv^*+\Nv_2}{\Nv_2}\right|\right]+
\mu\left[\frac{1}{2}\log\left|\frac{\Sv+\Nv_2}{\Bv^*+\Nv_2}\right|-
\frac{1}{2}\log\left|\frac{\Sv+\Nv_3}{\Bv^*+\Nv_3}\right|\right].\label{eq:c2}
\end{align}
Thus, we have obtained a contradiction between \eqref{eq:c1} and
\eqref{eq:c2}. As a result, all the achievable rate pairs must be
inside the secrecy rate region \eqref{eq:rr1a}. This completes the
proof of the theorem.

\section{Proof of Theorem~\ref{thm:DLMIMO2}} \label{sec:App2}
In this section, we prove Theorem~\ref{thm:DLMIMO2} following similar steps as
those used in the proof for Theorem~\ref{thm:DLMIMO1}. The achievability of the
secrecy rate region \eqref{eq:rr2a} can be obtained from the secrecy rate
region \eqref{eq:rr2} by letting $\Uv$ and $\Vv$ be two independent Gaussian
vectors with zero means and covariance matrices $\Sv-\Bv$ and $\Bv$,
respectively and $\Xv=\Uv+\Vv$. We therefore concentrate on the converse part
of the theorem.

To show that \eqref{eq:rr2a} is indeed the secrecy capacity region
of the vector Gaussian broadcast channel \eqref{eq:Ch}, we will use
proof by contradiction. Assume that $(R_1^o,R_2^o)$ is an achievable
secrecy rate pair that lies \emph{outside} the secrecy rate region
\eqref{eq:rr2a}. Note that $\Nv_1\preceq\Nv_3$. From
\cite[Theorem~1]{LS-ITSubm08}, we can bound $R_1^o$ by
\begin{align*}
R_1^o \le \frac{1}{2}\log\left|\frac{\Sv+\Nv_1}{\Nv_1}\right|-
\frac{1}{2}\log\left|\frac{\Sv+\Nv_3}{\Nv_3}\right| = R_1^{max}.
\end{align*}
Note that when $R_2^o=0$, $R_1^{max}$ is achievable by letting
$\Bv=\Sv$ in \eqref{eq:rr2}. Thus, we may assume that $R_2^o>0$ and
write $R_1^o=R_1^*+\delta$ for some $\delta>0$ where $R_1^*$ is
given by
\begin{align*}
\max_{\Bv} & \quad
\left[\frac{1}{2}\log\left|\frac{\Bv+\Nv_1}{\Nv_1}\right|-
\frac{1}{2}\log\left|\frac{\Bv+\Nv_3}{\Nv_3}\right|\right]\\
\mbox{subject to:}& \quad 0 \preceq \Bv \preceq \Sv\\
& \quad \frac{1}{2}\log\left|\frac{\Sv+\Nv_2}{\Bv+\Nv_2}\right|-
\frac{1}{2}\log\left|\frac{\Sv+\Nv_3}{\Bv+\Nv_3}\right| \ge R_2^o.
\end{align*}
Let $\Bv^*$ be an optimal solution to the above optimization
program. Then, $\Bv^*$ must satisfy the following KKT conditions:
\begin{align}
(\Bv^*+\Nv_1)^{-1}+(\mu-1)(\Bv^*+\Nv_3)^{-1}+\Mv_1&=\mu(\Bv^*+\Nv_2)^{-1}+\Mv_2\label{eq:kkt1a}\\
\Bv^*\Mv_1 & = 0\label{eq:kkt2a}\\
\mbox{and} \quad\quad (\Sv-\Bv^*)\Mv_2 & = 0\label{eq:kkt3a}
\end{align}
where $\Mv_1$ and $\Mv_2$ are $n \times n$ positive semidefinite
matrices, and $\mu$ is a nonnegative real scalar such that
$\mu\ge1$.\footnote{If $\mu<1$, it is easy to see that $\Bv^*=\Sv$
is an optimal solution and hence contradicts the assumption that
$R_2^o>0$.} Therefore,
$$R_2^o=\frac{1}{2}\log\left|\frac{\Sv+\Nv_2}{\Bv^*+\Nv_2}\right|-
\frac{1}{2}\log\left|\frac{\Sv+\Nv_3}{\Bv^*+\Nv_3}\right|$$ and
\begin{align}
R_1^o+\mu R_2^o =
\left[\frac{1}{2}\log\left|\frac{\Bv^*+\Nv_1}{\Nv_1}\right|-
\frac{1}{2}\log\left|\frac{\Bv^*+\Nv_3}{\Nv_3}\right|\right]+
\mu\left[\frac{1}{2}\log\left|\frac{\Sv+\Nv_2}{\Bv^*+\Nv_2}\right|-
\frac{1}{2}\log\left|\frac{\Sv+\Nv_3}{\Bv^*+\Nv_3}\right|\right]+\delta.
\label{eq:c1a}
\end{align}

On the other hand, by the converse part of Theorem~\ref{thm:dsbc2}
\begin{align}
R_1^o+\mu R_2^o \le\;&
[I(\Xv;\Xv+\Zv_1|U)-I(\Xv;\Xv+\Zv_3|U)]+\mu[I(U;\Xv+\Zv_2)-I(U;\Xv+\Zv_3)]\notag\\
=\;&[h(\Zv_3)-h(\Zv_1)]-\mu[h(\Xv+\Zv_3)-h(\Xv+\Zv_2)]+\notag\\
\;&[h(\Xv+\Zv_1|U)+(\mu-1)h(\Xv+\Zv_3|U)-\mu h(\Xv+\Zv_2|U)]\notag\\
\le\;&\frac{1}{2}\log\left|\frac{\Nv_3}{\Nv_1}\right|-\frac{\mu}{2}\log\left|\frac{\Sv+\Nv_3}{\Sv+\Nv_2}\right|+\notag\\
\;&[h(\Xv+\Zv_1|U)+(\mu-1)h(\Xv+\Zv_3|U)-\mu h(\Xv+\Zv_2|U)]
\label{eq:t5a}
\end{align}
for some jointly distributed $(U,\Xv)$ independent of
$(\Zv_1,\Zv_2,\Zv_3)$, where the last inequality follows from
\eqref{eq:t6}.

Since $\mu\ge1$, by letting
\begin{align*}
\mu_1=\frac{1}{\mu}, \quad \mu_3=\frac{\mu-1}{\mu}, \quad
\tilde{\Mv}_1=\frac{\Mv_1}{\mu}, \quad \mbox{and} \;\;
\tilde{\Mv}_2=\frac{\Mv_2}{\mu}
\end{align*}
we can rewrite the KKT conditions \eqref{eq:kkt1a}--\eqref{eq:kkt3a}
as
\begin{align*}
\mu_1(\Bv^*+\Nv_1)^{-1}+\mu_3(\Bv^*+\Nv_3)^{-1}+\tilde{\Mv}_1&=(\Bv^*+\Nv_2)^{-1}+\tilde{\Mv}_2\\
\Bv^*\tilde{\Mv}_1 & = 0\\
\mbox{and} \quad\quad (\Sv-\Bv^*)\tilde{\Mv}_2 & = 0.
\end{align*}
Thus, by Theorem~\ref{thm:LCVC-epi}
\begin{align}
h(\Xv+\Zv_1|U)+(\mu-1)h&(\Xv+\Zv_3|U)-\mu h(\Xv+\Zv_2|U)\notag\\
&\leq
\frac{1}{2}\log|\Bv^*+\Nv_1|+\frac{1-\mu}{2}\log|\Bv^*+\Nv_3|-\frac{\mu}{2}\log|\Bv^*+\Nv_2|.
\label{eq:t7a}
\end{align}
Substituting \eqref{eq:t7} into \eqref{eq:t5a}, we have
\begin{align}
R_1^o+\mu R_2^o \le&\;
\frac{1}{2}\log\left|\frac{\Nv_3}{\Nv_1}\right|-\frac{\mu}{2}\log\left|\frac{\Sv+\Nv_3}{\Sv+\Nv_2}\right|+\notag\\
&\;\left[\frac{1}{2}\log|\Bv^*+\Nv_1|+\frac{\mu-1}{2}\log|\Bv^*+\Nv_3|-\frac{\mu}{2}\log|\Bv^*+\Nv_2|\right]\notag\\
=&\;\left[\frac{1}{2}\log\left|\frac{\Bv^*+\Nv_1}{\Nv_1}\right|-
\frac{1}{2}\log\left|\frac{\Bv^*+\Nv_3}{\Nv_3}\right|\right]+
\mu\left[\frac{1}{2}\log\left|\frac{\Sv+\Nv_2}{\Bv^*+\Nv_2}\right|-
\frac{1}{2}\log\left|\frac{\Sv+\Nv_3}{\Bv^*+\Nv_3}\right|\right].\label{eq:c2a}
\end{align}
Thus, we have obtained a contradiction between \eqref{eq:c1a} and
\eqref{eq:c2a}. As a result, all the achievable rate pairs must be
inside the secrecy rate region \eqref{eq:rr2a}. This completes the
proof of the theorem.

\section{Conclusions} \label{sec:Con}
This paper has considered an EPI of Costa and has established a natural
generalization by replacing the scalar parameter in the original Costa EPI with
a matrix one. The generalized Costa EPI has been proven using a perturbation
approach via a fundamental relationship between the derivative of mutual
information and the MMSE in linear vector Gaussian channels. This is an example
of how the connections between information theory and statistics can be
explored to provide new mathematical tools for information theory.

As an application, a new extremal entropy inequality has been derived from the
generalized Costa EPI and then used to characterize the secrecy capacity
regions of the degraded vector Gaussian broadcast channel problem with layered
confidential messages. We expect that the generalized Costa EPI will also play
important roles in solving some other Gaussian network communication problems.

\appendices

\section{Proof of Theorem~\ref{thm:dsbc1}}\label{app:app}
\subsection{Achievability} We first show that the secrecy rate region
\eqref{eq:rr1} is achievable. Following the idea of superposition
coding for the degraded broadcast channel \cite{Ber-IT73}, we
introduce an auxiliary codebook which can be distinguished by both
receiver 1 and receiver 2. The codebook is generated using random
binning \cite{Wyn-BSTJ75,CK-IT78}.

Fix $p(u)$ and $p(x|u)$ and let
\begin{subequations} \label{eq:def-RR}
\begin{align}
&          & R'_1 &= I(X;Y_2|U)-\epsilon_1& \label{eq:def-RR-a}\\
&\text{and}& R'_2 &= I(U;Y_3)-\epsilon_1& \label{eq:def-RR-b}
\end{align}
\end{subequations}
for some $\epsilon_1 > 0$. Let
\begin{align*}
L_k &=2^{nR_k},\quad J_k=2^{nR'_k} \quad  \text{and}, \quad T_k=L_k
J_k  \quad k=1,2.
\end{align*}
Without loss of generality, $L_k$, $L'_k$ and $J_k$ are assumed to be integers.

\subsubsection*{Codebook generation} Generate $T_2$ independent
codewords $u^{n}$ of length $n$ according to $\prod_{i=1}^{n}p(u_i)$ and label
them as
$$u^{n}(w_2,j_2),\quad w_2 \in \{1,\dots,L_2\},\quad j_2 \in \{1,\dots, J_2\}.$$
For each codeword $u^{n}(w_2,j_2)$, generate $T_1$ independent codewords
$x^{n}$ according to $\prod_{i=1}^{n}p(x_i|u_i)$ and label them as
$$x^{n}(w_1,j_1,w_2,j_2)=x^{n}\bigl(w_1,j_1,u^{n}(w_2,j_2)\bigr),
\quad w_k \in \{1,\dots,L_k\} \quad \mbox{and} \quad j_k \in
\{1,\dots, J_k\}.$$

\subsubsection*{Encoding} To send a message pair $(w_1,w_2)$, the transmitter
randomly chooses a pair $(j_1,j_2)$ and sends the corresponding
codeword $x^{n}(w_1,j_1,w_2,j_2)$ through the channel.

\subsubsection*{Decoding} Receiver 2 determines the unique $w_2$ such that
$$\bigl(u^{n}(w_2,j_2), y_2^{n} \bigr)\in \mathcal{A}_{\epsilon}^{(n)}(p_{U,Y_2}) $$
where $\mathcal{A}_{\epsilon}^{(n)}(p_{U,Y_2})$ denotes the set of jointly
typical sequences $u^{n}$ and $y_2^{n}$ with respect to $p(u,y_2)$. If there
are none such or more than one such, an error is declared. Receiver 1 looks for
the unique $(w_1,w_2)$ such that
$$\bigl(u^{n}(w_2,j_2), x^{n}(w_1,j_1,w_2,j_2), y_1^{n} \bigr)\in \mathcal{A}_{\epsilon}^{(n)}(p_{U,X,Y_1})$$
where $\mathcal{A}_{\epsilon}^{(n)}(p_{U,X,Y_1})$ denotes the set of jointly
typical sequences $u^{n}$, $x^{n}$ and $y_1^{n}$ with respect to $p(u,x,y_1)$.
Otherwise, an error is declared.

\subsubsection*{Error probability analysis} By the symmetry of the
codebook generation, the probability error does not depend on which
codeword was sent. Hence, without loss of generality, we may assume
that the transmitter sends the message pair $(w_1,w_2)= (1,1)$
associated with the codeword $x^{n}(1,1,1,1)$ and define the
corresponding event
$$\mathcal{K}_1 := \{x^{n}(1,1,1,1) \; \text{was sent}\}.$$

First consider the decoding at receiver 2, for which we will show
that receiver 2 is able to decode $u^{n}(w_2,j_2)$ with small
probability of error if $R_2+R'_2<I(U; Y_2)$. To prove this, define
the event
\begin{align*}
\mathcal{E}_2(w_2,j_2):= \left\{\bigl(u^{n}(w_2,j_2), y_2^{n}
\bigr)\in \mathcal{A}_{\epsilon}^{(n)}(p_{U,Y_2})\right\}.
\end{align*}
Then, the probability of error at receiver 2 can be bounded from above as
\begin{align*}
P^{(n)}_{e,2}&\le \Pr \left\{\bigcap_{j_2}
\mathcal{E}_{2}^{c}(1,j_2)\Big|\mathcal{K}_{1}\right\}
+\sum_{w_2\neq 1,\,j_2} \Pr \{\mathcal{E}_{2}(w_2,j_2)|\mathcal{K}_{1}\}\notag\\
&\le \Pr \{\mathcal{E}_{2}^{c}(1,1)|\mathcal{K}_{1}\}+\sum_{w_2\neq
1,\,j_2} \Pr \{\mathcal{E}_{2}(w_2,j_2)|\mathcal{K}_{1}\}
\end{align*}
where
$$\mathcal{E}_{2}^{c}(1,j_2):=\left\{ \bigl(u^{n}(1,j_2), y_2^{n} \bigr)
\notin \mathcal{A}_{\epsilon}^{(n)}(p_{U,Y_2}) \right\}.$$ For large enough $n$
and $R_2+R_2'<I(U;Y_2)$, the joint asymptotic equipartition property (AEP)
\cite[Chapter~14.2]{Cover} implies
\begin{align}
P_{e,2}^{(n)}& \le \epsilon+T_2 2^{-n[I(U;Y_2)-\epsilon]} \notag\\
& =\epsilon+2^{n(R_2+R'_2)}\,2^{-n[I(U;Y_2)-\epsilon]} \notag \\
&\le 2 \epsilon. \label{eq:errb2}
\end{align}
Next, we will show that receiver 1 can successfully decode both
$u^{n}$ and $x^{n}$ if
\begin{align}
&           & R_1+R'_1&< I(X; Y_1|U)& \notag \\
&\text{and} & R_2+R'_2&< I(U; Y_2).&  \label{eq:dec-con}
\end{align}
Define the events
\begin{align*}
&   & \mathcal{E}_{1,1}(w_1,j_1,w_2,j_2) & := \left\{
\bigl(u^{n}(w_2,j_2), x^{n}(w_1,j_1,w_2,j_2), y_1^{n} \bigr)\in
\mathcal{A}_{\epsilon}^{(n)}(p_{U,X,Y_1}) \right\}.& \\
&\text{and} & \mathcal{E}_1(w_2,j_2)& := \left\{ \bigl(u^{n}(w_2,j_2), y_1^{n}
\bigr)\in \mathcal{A}_{\epsilon}^{(n)}(p_{U,Y_1}) \right\}&
\end{align*}
where $\mathcal{A}_{\epsilon}^{(n)}(p_{U,Y_1})$ denotes the set of jointly
typical sequences $u^{n}$ and $y_1^{n}$ with respect to $p(u,y_1)$. Then, the
probability of error
\begin{align*}
P^{(n)}_{e,1}&\le \Pr
\{\mathcal{E}_{1}^{c}(1,1)|\mathcal{K}_{1}\}+\sum_{w_2\neq 1,\,j_2}
\Pr \{\mathcal{E}_{1}(w_2,j_2)|\mathcal{K}_{1}\}+\sum_{w_1\neq
1,j_1,} \Pr \{\mathcal{E}_{1,1}(w_1,j_1,1,1)|\mathcal{K}_{1}\}
\end{align*}
where
$$\mathcal{E}_{1}^{c}(1,1):=\left\{ \bigl(u^{n}(1,1), y_1^{n} \bigr)
\notin \mathcal{A}_{\epsilon}^{(n)}(p_{U,Y_1})\right\}.$$ By the AEP
\cite[Chapter~14.2]{Cover},
\begin{align*}
&          & \Pr\{\mathcal{E}_{1}^{c}(1,1)|\mathcal{K}_{1}\} & \le \epsilon,&\\
&          & \Pr\{\mathcal{E}_{1}(w_2,j_2)|\mathcal{K}_{1}\}&\le 2^{-n[I(U;Y_1)-\epsilon]}, \quad \text{for} ~w_2\neq 1,&\\
&\text{and}&
\Pr\{\mathcal{E}_{1,1}(w_1,j_1,1,1)|\mathcal{K}_{1}\}&\le
2^{-n[I(X;Y_1|U)-\epsilon]}, \quad \text{for} ~w_1\neq 1.&
\end{align*}
Since the channel is degraded, we have $I(U;Y_1)\ge I(U;Y_2)$. Hence, if $n$ is
large enough and the condition (\ref{eq:dec-con}) holds, the probability of
error at receiver 1 can be bounded from above as
\begin{align}
P_{e,1}^{(n)}& \le \epsilon+T_2 2^{-n[I(U;Y_1)-\epsilon]}+T_1 2^{-n[I(X;Y_1|U)-\epsilon]}\notag\\
& \le \epsilon+2^{n(R_2+R'_2)}2^{-n[I(U;Y_2)-\epsilon]}+ 2^{n(R_1+R'_1)} 2^{-n[I(X;Y_1|U)-\epsilon]}\notag \\
&\le 3 \epsilon. \label{eq:errb1}
\end{align}
Together, (\ref{eq:errb2}) and (\ref{eq:errb1}) illustrate that
messages $(w_1,w_2)$ can be decoded at receiver 1 with a total
probability of error that goes to $0$ as long as the rate pair
$(R_1,R_2)$ satisfies \eqref{eq:rr1}.

\subsubsection*{Equivocation calculation} To show that (\ref{eq:eqv-r1}) holds, we
consider the following lower bound on the equivocation:
\begin{align}
H(W_1|Y_2^{n})& \ge H(W_1|Y_2^{n},U^{n})   \notag\\
&= H(W_1,Y_2^{n}|U^{n})-H(Y_2^{n}|U^{n})   \notag \\
&=
H(X^{n},Y_2^{n}|U^{n})-H(X^{n}|W_1,Y_2^{n},U^{n})-H(Y_2^{n}|U^{n}) \notag\\
&=
H(X^{n}|U^{n})+H(Y_2^{n}|X^{n},U^{n})-H(X^{n}|W_1,Y_2^{n},U^{n})-H(Y_2^{n}|U^{n}) \notag \\
&= H(X^{n}|U^{n})-H(X^{n}|W_1,Y_2^{n},U^{n})-I(X^{n};Y_2^{n}|U^{n})
\label{eq:ceq-2}
\end{align}
where the second equality is due to the fact that $W_1$ is
independent of everything else given $X^{n}$.

According to the codebook generation, for a given $U^{n}=u^{n}$,
$X^{n}$ has $T_1$ possible values with equal probabilities. Hence,
\begin{align}
 H(X^{n}|U^{n})&=n(R_1+R'_1)\notag\\
 &=n[R_1+I(X;Y_2|U)-\epsilon_1] \label{eq:ceq-3}
\end{align}
where (\ref{eq:ceq-3}) follows from the definition of $R'_1$ in
(\ref{eq:def-RR-a}).

Next, we show that for any given $\epsilon_2>0$, $H(X^{n}|W_1,Y_2^{n},U^{n})\le
n\epsilon_2$ for large enough $n$. To calculate $H(X^{n}|W_1,Y_2^{n},U^{n})$,
consider the following hypothetical scenario. Fix $W_1 = w_1$, and assume that
the transmitter sends a codeword $x^{n}\bigl(w_1,j_1,u^{n}(w_2,j_2)\bigr)$,
$j_1\in\{1,\dots,J_1\}$. Assume that receiver 2 knows the sequence $U^{n} =
u^{n}(w_2,j_2)$. Given index $W_1 = w_1$, receiver 2 decodes the codeword
$x^{n}(w_1,j_1,u^{n})$ (i.e., looks for the index $j_1$) based on the received
sequence $y_2$. Let $\lambda(w_1)$ denote the average probability of error of
decoding the index $j_1$ at receiver 2. By the AEP \cite[Chapter~14.2]{Cover},
we have $\lambda(w_1)\le \epsilon$ for sufficiently large $n$. By Fano's
inequality \cite[Chapter~2.11]{Cover},
\begin{align*}
\frac{1}{n} H(X^{n}|W_1=w_1,Y_2^{n},U^{n})
&\le \frac{1}{n}+\lambda(w_1)\frac{\log_2 J_1}{n}\notag\\
&\le \frac{1}{n}+\epsilon R'_1\notag\\
&:= \epsilon_2.
\end{align*}
Consequently,
\begin{align}
\frac{1}{n} H(X^{n}|W_1,Y_2^{n},U^{n})
&=\frac{1}{n}\sum_{w_1=1}^{L_1}\Pr(W_1=w_1) H(X^{n}|W_1=w_1,Y_2^{n},U^{n})\notag\\
&\le \epsilon_2. \label{eq:ceq-4}
\end{align}
By the AEP \cite[Chapter~14.2]{Cover}, for any $\epsilon_3$
\begin{align}
I(X^{n};Y_2^{n}|U^{n}) \le nI(X;Y_2|U)+ n\epsilon_3 \label{eq:ceq-5}
\end{align}
for sufficiently large $n$. Substituting (\ref{eq:ceq-3}),
(\ref{eq:ceq-4}) and (\ref{eq:ceq-5}) into (\ref{eq:ceq-2}), we have
\begin{align*}
\frac{1}{n}H(W_1|Y_2^{n})&\ge R_1-(\epsilon_1+\epsilon_2+\epsilon_3).
\end{align*}

Similarly, we can show that
\begin{align*}
H(W_2|Y_3^{n})\ge H(U^{n})-H(U^{n}|W_2,Y_3^{n})-I(U^{n};Y_3^{n})
\end{align*}
where
\begin{align*}
&    & &H(U^{n})  =n[R_2+I(U;Y_3)-\epsilon_1]&\\
&    & &H(U^{n}|W_2,Y_3^{n})  \le n\epsilon'_2&\\
&\text{and}& &I(U^{n};Y_3^{n})  \le n [ I(U;Y_3)+ \epsilon'_3],&
\end{align*}
where $\epsilon'_2$ and $\epsilon'_3$ vanishes in the limit as $n
\rightarrow \infty$. Hence,
\begin{align*}
\frac{1}{n}H(W_2|Y_3^{n})\ge R_2-(\epsilon_1+\epsilon'_2+\epsilon'_3).
\end{align*}
Note that $Y_3$ is degraded with respect to $Y_2$. Therefore,
\begin{eqnarray*}
H(W_1|Y_3^{n}) &\ge& H(W_1|Y_2^{n},Y_3^{n})\\
&=& H(W_1|Y_2^{n})\\
&\ge& R_1-(\epsilon_1+\epsilon_2+\epsilon_3).
\end{eqnarray*}
This proves the security condition (\ref{eq:eqv-r1}) and hence the
achievability part of the theorem.

\subsection{The Converse}
We first bound from above the secrecy rate $R_1$. The perfect
secrecy condition (\ref{eq:eqv-r1}) implies that for all
$\epsilon>0$,
\begin{subequations} \label{eq:eqv}
\begin{align}
&          & H(W_1|Y_2^{n})& \ge H(W_1)-n\epsilon & \label{eq:eqv1} \\
&\text{and}& H(W_2|Y_3^{n})& \ge H(W_2)-n\epsilon. & \label{eq:eqv2}
\end{align}
\end{subequations}
On the other hand, Fano's inequality \cite[Chapter~2.11]{Cover} implies that
for any $\epsilon_0>0$,
\begin{subequations} \label{eq:sd}
\begin{align}
&           & H(W_1|Y_1^{n}) &\le \epsilon_0 \log\left(2^{nR_1}-1\right)
            +h(\epsilon_0) := n\delta_1 & \label{eq:sd1}\\
&\text{and} & H(W_2|Y_2^{n}) &\le \epsilon_0 \log\left(2^{nR_2}-1\right)
            +h(\epsilon_0) := n\delta_2. &\label{eq:sd2}
\end{align}
\end{subequations}
Thus,
\begin{align}
nR_1 & = H(W_1) \notag \\
           &\le  \bigl[H(W_1|Y_2^{n})+n \epsilon\bigr]+ \bigl[n \delta_1-H(W_1|Y_1^{n})\bigr]\notag\\
           &\le  H(W_1,W_2|Y_2^{n})-H(W_1|Y_1^{n},W_2) + n (\epsilon+\delta_1) \notag\\
           &\le  H(W_1|Y_2^{n},W_2)-H(W_1|Y_1^{n},W_2) + n
           (\epsilon+\delta_1+\delta_2)  \label{eq:apout2}
\end{align}
where the first inequality follows from (\ref{eq:eqv1}) and (\ref{eq:sd1}), and
the last inequality follows from (\ref{eq:sd2}). Let
$\delta=\epsilon+\delta_1+\delta_2$. By the chain rule of the mutual
information \cite[Chapter~2.5]{Cover},
\begin{align}
n(R_1-\delta) &\le I(W_1;Y_1^{n}|W_2)-I(W_1;Y_2^{n}|W_2)\notag\\
              & =  \sum_{i=1}^{n}\left[ I(W_1;Y_{1,i}|W_2,Y_{1,i+1}^{n})
                        -I(W_1;Y_{2,i}|W_2,Y_2^{i-1})\right] \notag\\
& =  \sum_{i=1}^{n}\left[ I(W_1;Y_{1,i}|W_2,Y_{1,i+1}^{n},Y_2^{i-1})
                        -I(W_1;Y_{2,i}|W_2,Y_{1,i+1}^{n},Y_2^{i-1})\right] \label{eq:apout3}
\end{align}
where the last equality follows from \cite[Lemma~7]{CK-IT78}. Let
\begin{align}
V_i:= \left(Y_{1,i+1}^{n},Y_2^{i-1}\right). \label{eq:def-Ui}
\end{align}
We can further bound (\ref{eq:apout3}) from above as
\begin{align}
n(R_1-\delta) &\le \sum_{i=1}^{n} \left[ I(W_1,X_i;Y_{1,i}|W_2,V_i)
                        -I(W_1,X_i;Y_{2,i}|W_2,V_i)\right]\notag\\
&\qquad - \sum_{i=1}^{n}\left[ I(X_i;Y_{1,i}|W_1,W_2,V_i)-
             I(X_i;Y_{2,i}|W_1,W_2,V_i) \right]\notag\\
&\le \sum_{i=1}^{n} \left[ I(W_1,X_i;Y_{1,i}|W_2,V_i)
                        -I(W_1,X_i;Y_{2,i}|W_2,V_i)\right]\notag\\
& = \sum_{i=1}^{n} \left[ I(X_i;Y_{1,i}|W_2,V_i)
                        -I(X_i;Y_{2,i}|W_2,V_i)\right]\label{eq:apout5}
\end{align}
where the second inequality follows from the Markov relation
$$(W_1,W_2,V_i)\rightarrow X_i \rightarrow Y_{1,i} \rightarrow
Y_{2,i},$$ and the last equality is due to the fact that $Y_{1,i}$ and
$Y_{2,i}$ are conditionally independent of everything else given $X_i$.

Next, we bound from above the secrecy rate $R_2$. By (\ref{eq:eqv2})
and (\ref{eq:sd2}),
\begin{align}
nR_2 & = H(W_2) \notag \\
     &\le  \bigl[H(W_2|Y_3^{n})+n \epsilon\bigr]+
           \bigl[n \delta_2-H(W_2|Y_2^{n})\bigr] \notag \\
     &=I(W_2;Y_2^{n})-I(W_2;Y_3^{n})+n (\epsilon+\delta_2) \notag\\
     &=\sum_{i=1}^{n} \left[ I(W_2;Y_{2,i}|Y_{2,i+1}^{n})
      -I(W_2;Y_{3,i}|Y_3^{i-1})\right]+n (\epsilon+\delta_2).
     \label{eq:apout7}
\end{align}
Let $\delta'  := \epsilon+\delta_2$ and
\begin{align}
V'_i & := \left(Y_{2,i+1}^{n},Y_3^{i-1}\right).  \label{eq:def-U2}
\end{align}
Applying \cite[Lemma~7]{CK-IT78} again, we may obtain
\begin{align}
n(R_2-\delta') &\le \sum_{i=1}^{n} \left[ I(W_2;Y_{2,i}|V'_i)
                     -I(W_2;Y_{3,i}|V'_i)\right] \notag\\
       &= \sum_{i=1}^{n} \left[ I(W_2,V'_i;Y_{2,i})
                  -I(W_2,V'_i;Y_{3,i})\right]- \sum_{i=1}^{n} \left[ I(V'_i;Y_{2,i})
                  -I(V'_i;Y_{3,i})\right] \notag \\
      & \le  \sum_{i=1}^{n} \left[ I(W_2,V'_i;Y_{2,i})
      -I(W_2,V'_i;Y_{3,i})\right] \label{eq:apout8}
\end{align}
where the last inequality follows from the Markov relation $V'_i
\rightarrow Y_{1,i} \rightarrow Y_{2,i}$. Furthermore, by the
definitions of $V_i$ and $V_i'$ in (\ref{eq:def-Ui}) and
(\ref{eq:def-U2}) respectively,
\begin{align}
V'_i \rightarrow (W_2,V_i) \rightarrow (Y_{2,i},Y_{3,i}).
\label{eq:apout8a}
\end{align}
By (\ref{eq:apout8}) and (\ref{eq:apout8a}),
\begin{align}
n(R_2-\delta') & \le  \sum_{i=1}^{n} \left[ I(W_2,V'_i,V_i;Y_{2,i})
      -I(W_2,V'_i,V_i;Y_{3,i})\right]-\sum_{i=1}^{n} \left[ I(V_i;Y_{2,i}|W_2,V'_i)
      -I(V_i;Y_{3,i}|W_2,V'_i)\right] \notag \\
      & = \sum_{i=1}^{n} \left[ I(W_2,V_i;Y_{2,i})
      -I(W_2,V_i;Y_{3,i})\right]-\sum_{i=1}^{n} \left[ I(V_i;Y_{2,i}|W_2,V'_i)
      -I(V_i;Y_{3,i}|W_2,V'_i)\right].\label{eq:apout9}
\end{align}
Note that $Y_{3,i}$ is conditionally independent of everything else given
$Y_{2,i}$. Hence,
\begin{align}
I(V_i;Y_{3,i}|W_2,V'_i) &\le I(V_i;Y_{2,i},Y_{3,i}|W_2,V'_i) \notag\\
    &= I(V_i;Y_{2,i}|W_2,V'_i)+I(V_i;Y_{3,i}|Y_{2,i},W_2,V'_i)\notag\\
    &= I(V_i;Y_{2,i}|W_2,V'_i). \label{eq:app-mo1}
\end{align}
Substituting (\ref{eq:app-mo1}) into (\ref{eq:apout9}), we have
\begin{align}
R_2& \le \frac{1}{n}\sum_{i=1}^{n} \left[ I(W_2, V_i;Y_{2,i})
      -I(W_2,V_i;Y_{3,i})\right]+\delta' .\label{eq:apupr2}
\end{align}

Finally, let
\begin{align}
U_i:= (W_2,V_i). \label{eq:def-V}
\end{align}
With this definition, (\ref{eq:apout5}) and \eqref{eq:apupr2} can be
rewritten as
\begin{align}
R_1 & \le \frac{1}{n}\sum_{i=1}^{n} \left[ I(X_i;Y_{1,i}|U_i)
      -I(X_i;Y_{2,i}|U_i)\right]+\delta. \notag \\
\mbox{and} \quad \quad R_2 & \le \frac{1}{n}\sum_{i=1}^{n} \left[
I(U_i;Y_{2,i})
      -I(U_i;Y_{3,i})\right]+\delta'.
\end{align}
Following the standard single-letterization process (e.g., see
\cite[Chapter~14.3]{Cover}), we have the desired converse result.

\section{Proof of Corollary~\ref{cor:cond}}
\label{app:cond} Fix $U=u$. By the generalized Costa EPI \eqref{eq:VC-epi}, we
have
\begin{align}
h(\Xv+\Av^{\frac{1}{2}}\Zv|U=u) &\ge \frac{n}{2}\log\left\{
|\Iv-\Av|^{\frac{1}{n}}\exp\left[\frac{2}{n}h(\Xv|U=u)\right]+
|\Av|^{\frac{1}{n}}\exp\left[\frac{2}{n}h(\Xv+\Zv|U=u)\right]\right\}.
\label{eq:cod-epi}
\end{align}
Taking expectation over $U$ on both sides of \eqref{eq:cod-epi}, we
may obtain
\begin{align}
h(\Xv+\Av^{\frac{1}{2}}\Zv|U) &\ge \frac{n}{2}\E\left[\log\left\{
|\Iv-\Av|^{\frac{1}{n}}\exp\left[\frac{2}{n}h(\Xv|U=u)\right]+
|\Av|^{\frac{1}{n}}\exp\left[\frac{2}{n}h(\Xv+\Zv|U=u)\right]\right\}\right] \notag\\
&\ge \frac{n}{2}\log \left\{
|\Iv-\Av|^{\frac{1}{n}}\exp\left[\frac{2}{n}\E\left[
h(\Xv|U=u)\right]\right]+ |\Av|^{\frac{1}{n}}\exp\left[\frac{2}{n}\E
\left[
h(\Xv+\Zv|U=u)\right]\right]\right\}\notag\\
&= \frac{n}{2}\log \left\{
|\Iv-\Av|^{\frac{1}{n}}\exp\left[\frac{2}{n} h(\Xv|U)\right]+
|\Av|^{\frac{1}{n}}\exp\left[\frac{2}{n} h(\Xv+\Zv|U)\right]\right\}
\label{eq:jen}
\end{align}
where the second inequality follows from Jensen's inequality
\cite[Chapter~2.6]{Cover} and the convexity of
$\log\left(a_1e^{x_1}+a_2e^{x_2}\right)$ in $(x_1,x_2)$ for $a_1,a_2\geq0$.
Taking logarithm on both sides of \eqref{eq:jen} proves the desired inequality
\eqref{eq:CVC-epi}.

\section{Proof of Corollary~\ref{cor:Lepi-2}}
\label{app:Lepi-2} Note that when $\mu=0$, \eqref{eq:LVC-2} implies
that $\Nv_1=\Nv_2$. Thus, both sides of \eqref{eq:Lepi-2} are equal
to zero and the inequality holds trivially with an equality. For the
rest of the proof, we will assume that $\mu>0$. The proof is rather
long so we divide it into several steps.

\emph{Step 1--Generalized eigenvalue decomposition.} \label{sec:GED}
We start by applying generalized eigenvalue decomposition
\cite{Gilbert} to the positive define matrices $\Bv^*+\Nv_1$ and
$\Bv^*+\Nv_2$. There exists an \emph{invertible} generalized
eigenvector matrix $\Vv$ such that
\begin{align}
          &  \Vv^{\T}(\Bv^*+\Nv_1)\Vv=\mathbf{\Lambda}_1\\
\text{and} \quad \quad & \Vv^{\T}(\Bv^*+\Nv_2)\Vv=\mathbf{\Lambda}_2
\label{eq:def-L12}
\end{align}
where $\mathbf{\Lambda}_1$ and $\mathbf{\Lambda}_2$ are positive
definite \emph{diagonal} matrices. Let
\begin{align}
\mathbf{\Lambda}_3:=\Vv^{\T}(\Bv^*+\Nv_3)\Vv \label{eq:def-L3}
\end{align}
be an $n \times n$ positive definite matrix. By \eqref{eq:LVC-2},
\begin{align}
\mathbf{\Lambda}_1^{-1}+\mu\mathbf{\Lambda}_3^{-1}=(1+\mu)
\mathbf{\Lambda}_2^{-1}. \label{eq:LV-cd2}
\end{align}
Thus, $\mathbf{\Lambda}_3$ is also diagonal. Moreover, since $\Nv_1
\preceq \Nv_3$,
\begin{align*}
\mathbf{\Lambda}_1-\mathbf{\Lambda}_3 =\Vv^\T(\Nv_1-\Nv_3)\Vv
\preceq 0.
\end{align*}
and hence
\begin{align}
\mathbf{\Lambda}_1 \preceq \mathbf{\Lambda}_3. \label{eq:ord1}
\end{align}

\emph{Step 2--Choosing matrix parameter $\Av$.} Let
$\tilde{\mathbf{\Lambda}}_3=\mathbf{\Lambda}_3+\epsilon\Iv$ for some
$\epsilon>0$, and let $\tilde{\mathbf{\Lambda}}_2$ be an $n \times
n$ matrix such that
\begin{align}
\mathbf{\Lambda}_1^{-1}+\mu\tilde{\mathbf{\Lambda}}_3^{-1}=(1+\mu)
\tilde{\mathbf{\Lambda}}_2^{-1}. \label{eq:LV-cd3}
\end{align}
Clearly, $\tilde{\mathbf{\Lambda}}_2$ is diagonal. Moreover, by
\eqref{eq:ord1}
\begin{align}
\mathbf{\Lambda}_1 \prec \tilde{\mathbf{\Lambda}}_3. \label{eq:ord0}
\end{align}
Note that $\mu>0$ so by \eqref{eq:LV-cd3} and \eqref{eq:ord0}
\begin{align}
\mathbf{\Lambda}_1 \prec \tilde{\mathbf{\Lambda}}_2 \prec
\tilde{\mathbf{\Lambda}}_3. \label{eq:ord2}
\end{align}
Comparing \eqref{eq:LV-cd2} and \eqref{eq:LV-cd3} and using the fact
that $\mathbf{\Lambda}_3 \prec \tilde{\mathbf{\Lambda}}_3$, we have
\begin{align}
\mathbf{\Lambda}_2 \prec \tilde{\mathbf{\Lambda}}_2. \label{eq:ord3}
\end{align}
Now let
\begin{align*}
& \Yv_1 := \Vv^\T(\Xv+\Zv_1)\\
& \Yv_2 := \Vv^\T(\Xv+\Zt_2)\\
\mbox{and} \quad \quad & \Yv_3 := \Vv^\T(\Xv+\Zt_3)
\end{align*}
where $\Zt_2$ and $\Zt_3$ are Gaussian $n$-vectors with covariance
matrices
\begin{align*}
\Nt_2 &= \Vv^{-\T}\tilde{\mathbf{\Lambda}}_2\Vv^{-1}-\Bv^*\\
&\succ \Vv^{-\T}\mathbf{\Lambda}_2\Vv^{-1}-\Bv^*\\
&=(\Bv^*+\Nv_2)-\Bv^*\\
&=\Nv_2
\end{align*}
and
\begin{align*}
\Nt_3 &= \Vv^{-\T}\tilde{\mathbf{\Lambda}}_3\Vv^{-1}-\Bv^*\\
&= \Vv^{-\T}(\mathbf{\Lambda}_3+\epsilon\Iv)\Vv^{-1}-\Bv^*\\
&=(\Bv^*+\Nv_3+\epsilon\Vv^{-\T}\Vv^{-1})-\Bv^*\\
&=\Nv_3+\epsilon\Vv^{-\T}\Vv^{-1}
\end{align*}
respectively and are independent of $\Xv$. The covariance matrices of $\Yv_k$,
$k=1,2,3$, can be calculated as
$\Vv^\T[\Cov(\Xv)-\Bv^*]\Vv+\mathbf{\Lambda}_1$,
$\Vv^\T[\Cov(\Xv)-\Bv^*]\Vv+\tilde{\mathbf{\Lambda}}_2$ and
$\Vv^\T[\Cov(\Xv)-\Bv^*]\Vv+\tilde{\mathbf{\Lambda}}_3$, respectively. Thus,
$\Yv_2$ and $\Yv_3$ can be equivalently written as
\begin{align*}
& \Yv_3 = \Yv_1+\Zv\\
\mbox{and} \quad \quad & \Yv_2=\Yv_1+\Av^{\frac{1}{2}}\Zv
\end{align*}
where $\Zv$ is a Gaussian $n$-vector with covariance matrix
$\tilde{\mathbf{\Lambda}}_3-\mathbf{\Lambda}_1\succ0$ and is
independent of $\Yv_1$, and
\begin{align}
\Av&:=(\tilde{\mathbf{\Lambda}}_2-\mathbf{\Lambda}_1)(\tilde{\mathbf{\Lambda}}_3-\mathbf{\Lambda}_1)^{-1}.
\label{eq:def-A}
\end{align}
Clearly, $\Av$ is diagonal. Moreover, by \eqref{eq:ord2} $0 \prec
\Av \prec \Iv$.

\emph{Step 3--Applying generalized Costa's EPI.} By the generalized Costa EPI
\eqref{eq:VC-epi},
\begin{align*}
h(\Yv_2|U)\ge
\frac{n}{2}\log\left\{|\Iv-\Av|^{\frac{1}{n}}\exp\left[\frac{2}{n}h(\Yv_1|U)\right]
+|\Av|^{\frac{1}{n}}\exp\left[\frac{2}{n}h(\Yv_3|U)\right]\right\}.
\end{align*}
Thus,
\begin{align}
h(&\Yv_1|U)+\mu h(\Yv_3|U) -(1+\mu)h(\Yv_2|U) \notag\\
&\le h(\Yv_1|U)+\mu h(\Yv_3|U) -
\frac{(1+\mu)n}{2}\log\left\{|\Iv-\Av|^{\frac{1}{n}}\exp\left[\frac{2}{n}h(\Yv_1|U)\right]
+|\Av|^{\frac{1}{n}}\exp\left[\frac{2}{n}h(\Yv_3|U)\right]\right\}.\label{eq:hhL}
\end{align}
Now we consider the function
\begin{align*}
f(b,c)=b+\mu c
-\frac{(1+\mu)n}{2}\log\left[|\Iv-\Av|^{\frac{1}{n}}\exp\left(\frac{2b}{n}\right)
+|\Av|^{\frac{1}{n}}\exp\left(\frac{2c}{n}\right)\right].
\end{align*}
Note that
\begin{align*}
\nabla f(b,c) &= \left[\begin{matrix} \displaystyle{1-(1+\mu)
\frac{|\Iv-\Av|^{\frac{1}{n}}\exp(2b/n)}{|\Iv-\Av|^{\frac{1}{n}}\exp(2b/n)+|\Av|^{\frac{1}{n}}\exp(2c/n)}
}\\[3mm]
\displaystyle{\mu-(1+\mu)
\frac{|\Av|^{\frac{1}{n}}\exp(2c/n)}{|\Iv-\Av|^{\frac{1}{n}}\exp(2b/n)+|\Av|^{\frac{1}{n}}\exp(2c/n)}}
\end{matrix}\right]
\end{align*}
and
\begin{align*}
\nabla^{2}f(b,c)=
-\frac{2(1+\mu)}{n}\frac{|\Av|^{\frac{1}{n}}|\Iv-\Av|^{\frac{1}{n}}
\exp[(2b+2c)/n]}{\left[|\Iv-\Av|^{\frac{1}{n}}
\exp(2b/n)+|\Av|^{\frac{1}{n}} \exp(2c/n)\right]^2}
\left[\begin{matrix}1 & -1 \\-1 & 1\end{matrix}\right] \preceq 0.
\end{align*}
So $f(b,c)$ is concave in $(b,c)$. By setting $\nabla f(b,c)=0$, the
global maximum is achieved when
\begin{align*}
c=b+\frac{n}{2}\log \left[\mu
\left(\frac{|\Iv-\Av|}{|\Av|}\right)^{\frac{1}{n}}\right]
\end{align*}
and the maximum is given by
\begin{align*}
\frac{\mu
n}{2}\log\left[\mu\left(\frac{|\Iv-\Av|}{|\Av|}\right)^{\frac{1}{n}}\right]
-\frac{(1+\mu)n}{2}\log\left[(1+\mu)|\Iv-\Av|^{\frac{1}{n}}\right].
\end{align*}
Hence,
\begin{align}
h(\Yv_1|U)+&\mu h(\Yv_3|U)-(1+\mu)h(\Yv_2|U)\notag\\
&\le\frac{\mu n}{2}\log\left[\mu
 \left(\frac{|\Iv-\Av|}{|\Av|}\right)^{\frac{1}{n}}\right]
 -\frac{(1+\mu)n}{2}\log\left[(1+\mu)|\Iv-\Av|^{\frac{1}{n}}\right].
 \label{eq:h123}
\end{align}

\emph{Step 4--Calculating $\log|\Av|$ and $\log|\Iv-\Av|$.} Note that
\eqref{eq:LV-cd3} can be rewritten as
\begin{align*}
\mu(\mathbf{\Lambda}_1^{-1}-\tilde{\mathbf{\Lambda}}_3^{-1})
=(1+\mu)(\mathbf{\Lambda}_1^{-1}-\tilde{\mathbf{\Lambda}}_2^{-1})
\end{align*}
which gives
\begin{align}
\left|\frac{\tilde{\mathbf{\Lambda}}_2-\mathbf{\Lambda}_1}
{\tilde{\mathbf{\Lambda}}_3-\mathbf{\Lambda}_1}\right|
&=\left(\frac{\mu}{1+\mu}\right)^n
\left|\frac{\tilde{\mathbf{\Lambda}}_2}{\tilde{\mathbf{\Lambda}}_3}\right|.
\label{eq:equ1}
\end{align}
Similarly, we have
\begin{align*}
(\mathbf{\Lambda}_1^{-1}-\tilde{\mathbf{\Lambda}}_3^{-1})
=(1+\mu)(\tilde{\mathbf{\Lambda}}_2^{-1}-\tilde{\mathbf{\Lambda}}_3^{-1})
\end{align*}
and hence
\begin{align}
\left|\frac{\tilde{\mathbf{\Lambda}}_3-\tilde{\mathbf{\Lambda}}_2}
{\tilde{\mathbf{\Lambda}}_3-\mathbf{\Lambda}_1}\right|
&=\left(\frac{1}{1+\mu}\right)^n
\left|\frac{\tilde{\mathbf{\Lambda}}_2}{\tilde{\mathbf{\Lambda}}_1}\right|.
\label{eq:equ2}
\end{align}
According to the definition of $\Av$ in \eqref{eq:def-A},
\begin{align}
\log|\Av| &=
\log\left|\frac{\tilde{\mathbf{\Lambda}}_2-\mathbf{\Lambda}_1}
{\tilde{\mathbf{\Lambda}}_3-\mathbf{\Lambda}_1}\right|\notag\\
&=\log\left[\left(\frac{\mu}{1+\mu}\right)^n
\left|\frac{\tilde{\mathbf{\Lambda}}_2}{\tilde{\mathbf{\Lambda}}_3}\right|\right]
\label{eq:equ3}
\end{align}
and
\begin{align}
\log|\Iv-\Av| &=
\log\left|\frac{\tilde{\mathbf{\Lambda}}_3-\tilde{\mathbf{\Lambda}}_2}
{\tilde{\mathbf{\Lambda}}_3-\mathbf{\Lambda}_1}\right|\notag\\
&=\log\left[\left(\frac{1}{1+\mu}\right)^n
\left|\frac{\tilde{\mathbf{\Lambda}}_2}{\mathbf{\Lambda}_1}\right|\right]
\label{eq:equ4}
\end{align}
where \eqref{eq:equ3} and \eqref{eq:equ4} follow \eqref{eq:equ1} and
\eqref{eq:equ2}, respectively. Substituting \eqref{eq:equ3} and
\eqref{eq:equ4} into \eqref{eq:h123}, we have
\begin{align}
h(\Yv_1|U)+\mu
h(\Yv_3|U)-(1+\mu)h(\Yv_2|U)&\le\frac{1}{2}\log|\mathbf{\Lambda}_1|+
\frac{\mu}{2}\log|\tilde{\mathbf{\Lambda}}_3|-
\frac{1+\mu}{2}\log|\tilde{\mathbf{\Lambda}}_2|. \label{eq:h123b}
\end{align}

\emph{Step 5--Letting $\epsilon \downarrow 0$.} Note that
$\tilde{\mathbf{\Lambda}}_3=\mathbf{\Lambda}_3+\epsilon\Iv
\rightarrow \mathbf{\Lambda}_3$ and
$\Nt_3=\Nv_3+\epsilon\Vv^{-\T}\Vv^{-1} \rightarrow \Nv_3$ in the
limit as $\epsilon \downarrow 0$. Moreover, by \eqref{eq:LV-cd3} we
have $\tilde{\mathbf{\Lambda}}_2 \rightarrow \mathbf{\Lambda}_2$ and
hence
\begin{align*}
\Nt_2 &= \Vv^{-\T}\tilde{\mathbf{\Lambda}}_2\Vv^{-1}-\Bv^*\\
& \rightarrow \Vv^{-\T}\mathbf{\Lambda}_2\Vv^{-1}-\Bv^*\\
&=(\Bv^*+\Nv_2)-\Bv^*\\
&=\Nv_2.
\end{align*}
Letting $\epsilon \downarrow 0$ on both sides of \eqref{eq:h123b},
we have
\begin{align}
h(\Vv^\T(\Xv+\Nv_1)|U)+\mu h(\Vv^\T(\Xv+\Nv_3)|U)-&(1+\mu)h(\Vv^\T(\Xv+\Nv_2)|U)\notag\\
&\le\frac{1}{2}\log|\mathbf{\Lambda}_1|+
\frac{\mu}{2}\log|\mathbf{\Lambda}_3|-
\frac{1+\mu}{2}\log|\mathbf{\Lambda}_2|. \label{eq:h123c}
\end{align}
Using the fact that
\begin{align*}
h(\Vv^\T(\Xv+\Nv_1)|U)=h(\Xv+\Nv_1|U)+\log|\Vv|
\end{align*}
and
\begin{align*}
\log|\mathbf{\Lambda}_k|&=\log|\Vv^{\T}(\Bv^*+\Nv_k)\Vv|\\
&=\log|\Bv^*+\Nv_k|+2\log|\Vv|
\end{align*}
for $k=1,2,3$, the desired inequality \eqref{eq:Lepi-2} can be
obtained from \eqref{eq:h123c}. This completes the proof of the
corollary.

\section{Proof of Corollary~\ref{cor:Lepi-K}} \label{app:Lepi-K}
Here, we prove Corollary~\ref{cor:Lepi-K} using mathematical
induction. Note that when $K=1$, \eqref{eq:LVC-K} implies that
$\Nv_1=\Nv_0$. Thus, the inequality \eqref{eq:Lepi-K} holds
trivially with equality for any $(U,\Xv)$ independent of
$(\Zv_0,\Zv_1)$.

Assume that the inequality \eqref{eq:Lepi-K} holds for $K=Q-1$. Let
$\Nv$ be an $n \times n$ symmetric matrix such that
\begin{align}
(\Bv^*+\Nv)^{-1}=\sum_{k=1}^{Q-1}\mu_k'(\Bv^*+\Nv_k)^{-1}
\label{eq:Nt3-1}
\end{align}
where
\begin{align*}
\mu_k':=\frac{\mu_k}{\sum_{j=1}^{Q-1}\mu_j}, \quad j=1,\ldots,Q.
\end{align*}
By the assumption $\Nv_1 \preceq \ldots \preceq \Nv_{Q-1}$, we have
from \eqref{eq:Nt3-1}
\begin{align}
\Nv_1 \preceq \Nv \preceq \Nv_{Q-1}. \label{eq:ord4}
\end{align}
Let $\Zv$ be a Gaussian random $n$-vector with covariance matrix
$\Nv$ and independent of $(U,\Xv)$. By the induction assumption and
\eqref{eq:Nt3-1},
\begin{align}
\sum_{k=1}^{Q-1} \mu_k' h(\Xv+\Zv_k|U)-h(\Xv+\Zv|U)&\leq
\sum_{k=1}^{Q-1} \frac{\mu_k'}{2} \log|\Bv+\Nv_k|
-\frac{1}{2}\log|\Bv+\Nv|.\label{eq:Qv-3}
\end{align}

On the other hand, substitute \eqref{eq:Nt3-1} into \eqref{eq:LVC-K}
and we have
\begin{align*}
(\Bv+\Nv)^{-1}+\mu_Q'(\Bv+\Nv_Q)^{-1}=(1+\mu_Q')(\Bv+\Nv_0)^{-1}.
\end{align*}
Note from \eqref{eq:ord4} that $\Nv \preceq \Nv_{Q-1} \preceq
\Nv_Q$. Thus, by Corollary~\ref{cor:Lepi-2}
\begin{align}
h(\Xv+\Zv|U) +  \mu_Q' h&(\Xv+\Zv_Q|U) - (1+\mu_Q')h(\Xv+\Zv_0|U)\notag\\
&\leq \frac{1}{2}\log|\Bv+\Nv|+
\frac{\mu_Q'}{2}\log|\Bv+\Nv_Q|-\frac{1+\mu_Q'}{2}\log|\Bv+\Nv_0|.
\label{eq:Qv-4}
\end{align}
Putting together \eqref{eq:Qv-3} and \eqref{eq:Qv-4}, we have
\begin{align*}
\sum_{j=1}^{Q} \mu_j h(\Xv+\Zv_j|U)- h(\Xv+\Zv_0|U)&\leq
\sum_{j=1}^{Q} \frac{\mu_j}{2} \log|\Bv+\Nv_j|
-\frac{1}{2}\log|\Bv+\Nv_0|.
\end{align*}
This proved the induction step and hence the corollary.

\bibliographystyle{IEEEtran}
\bibliography{secrecy}

\end{document}